\documentclass[copyright,creativecommons]{eptcs} 

\usepackage{iftex}
\usepackage{tikzit}
\usepackage{amsfonts, amsmath, amssymb}
\usepackage{braket}
\usepackage{colortbl}
\usepackage{algorithm}
\usepackage{algpseudocode}
\usepackage{listings}
\usepackage{subcaption}

\ifpdf
  \usepackage{underscore}         
  \usepackage[T1]{fontenc}        
\else
  \usepackage{breakurl}           
\fi

\tikzstyle{gate}=[shape=rectangle, text height=1.5ex, text depth=0.25ex, yshift=0.5mm, fill=white, draw=black, minimum height=5mm, yshift=-0.5mm, minimum width=5mm, font={\small}, tikzit category=circuit]
\tikzstyle{big gate}=[shape=rectangle, text height=1.5ex, text depth=0.25ex, yshift=0.5mm, fill=white, draw=black, minimum height=10mm, yshift=-0.5mm, minimum width=5mm, font={\small}, tikzit category=circuit]
\tikzstyle{Z dot}=[inner sep=0mm, minimum size=2mm, shape=circle, draw=black, fill={rgb,255: red,221; green,255; blue,221}, tikzit category=zx]
\tikzstyle{Z phase dot}=[minimum size=5mm, font={\footnotesize\boldmath}, shape=rectangle, rounded corners=2mm, inner sep=0.2mm, outer sep=-2mm, scale=0.8, tikzit shape=rectangle, draw=black, fill={rgb,255: red,221; green,255; blue,221}, tikzit draw=blue, tikzit category=zx]
\tikzstyle{X dot}=[Z dot, shape=circle, draw=black, fill={rgb,255: red,255; green,136; blue,136}, tikzit category=zx]
\tikzstyle{X phase dot}=[Z phase dot, tikzit shape=rectangle, tikzit draw=blue, fill={rgb,255: red,255; green,136; blue,136}, font={\footnotesize\boldmath}, tikzit category=zx]
\tikzstyle{hadamard}=[fill=yellow, draw=black, shape=rectangle, inner sep=0.6mm, minimum height=1.5mm, minimum width=1.5mm, tikzit category=zx]
\tikzstyle{paulibox}=[fill={rgb,255: red,221; green,221; blue,255}, draw=black, shape=rectangle, inner sep=0.6mm, minimum height=5mm, minimum width=5mm, font={\footnotesize}, text height=1.5ex, text depth=0.25ex, tikzit category=zx]
\tikzstyle{vertex}=[inner sep=0mm, minimum size=1mm, shape=circle, draw=black, fill=black, tikzit category=misc]
\tikzstyle{vertex set}=[inner sep=0mm, minimum size=1mm, shape=circle, draw=black, fill=white, font={\footnotesize\boldmath}, tikzit category=misc]
\tikzstyle{small black dot}=[fill=black, draw=black, shape=circle, inner sep=0pt, minimum width=1.2mm, tikzit category=circuit]
\tikzstyle{cnot ctrl}=[fill=black, draw=black, shape=circle, inner sep=0pt, minimum width=1.2mm, tikzit category=circuit]
\tikzstyle{cnot targ}=[fill=white, draw=white, shape=circle, tikzit category=circuit, label={center:$\oplus$}, inner sep=0pt, minimum width=2.1mm, tikzit fill={rgb,255: red,102; green,204; blue,255}, tikzit draw=black]
\tikzstyle{ket}=[fill=white, draw=black, shape=regular polygon, regular polygon sides=3, regular polygon rotate=-30, scale=0.7, inner sep=1pt, tikzit category=circuit, tikzit shape=rectangle, tikzit fill=green]
\tikzstyle{bra}=[fill=white, draw=black, shape=regular polygon, regular polygon sides=3, regular polygon rotate=30, scale=0.7, inner sep=1pt, tikzit category=circuit, tikzit shape=rectangle, tikzit fill=red]
\tikzstyle{scalar}=[shape=rectangle, text height=1.5ex, text depth=0.25ex, yshift=0.5mm, fill=white, draw=black, minimum height=5mm, yshift=-0.5mm, minimum width=5mm, font={\small}, rounded corners=2mm]
\tikzstyle{clabel}=[fill=white, draw=none, shape=rectangle, tikzit fill={rgb,255: red,56; green,255; blue,242}, font={\footnotesize}, inner sep=1pt, tikzit category=labels]
\tikzstyle{empty diagram}=[draw={gray!40!white}, dashed, shape=rectangle, minimum width=1cm, minimum height=1cm, tikzit category=misc]
\tikzstyle{amap}=[fill=white, draw=black, shape=NEbox, tikzit category=asymmetric, tikzit fill=yellow, tikzit shape=rectangle]
\tikzstyle{amap conj}=[fill=white, draw=black, shape=NWbox, tikzit category=asymmetric, tikzit fill=green, tikzit shape=rectangle]
\tikzstyle{amap adj}=[fill=white, draw=black, shape=SEbox, tikzit category=asymmetric, tikzit fill=red, tikzit shape=rectangle]
\tikzstyle{amap trans}=[fill=white, draw=black, shape=SWbox, tikzit category=asymmetric, tikzit fill=orange, tikzit shape=rectangle]
\tikzstyle{astate}=[fill=white, draw=black, shape=NEtriangle, tikzit category=asymmetric, tikzit shape=circle, tikzit fill=yellow]
\tikzstyle{astate conj}=[fill=white, draw=black, shape=NWtriangle, tikzit category=asymmetric, tikzit shape=circle, tikzit fill=green]
\tikzstyle{astate adj}=[fill=white, draw=black, shape=SEtriangle, tikzit category=asymmetric, tikzit shape=circle, tikzit fill=red]
\tikzstyle{astate trans}=[fill=white, draw=black, shape=SWtriangle, tikzit category=asymmetric, tikzit shape=circle, tikzit fill=orange]
\tikzstyle{bigbox}=[fill=white, draw=black, shape=rectangle, tikzit category=zx, minimum width=1.5cm, minimum height=2.6cm]
\tikzstyle{bigger gate}=[shape=rectangle, text height=1.5ex, text depth=0.25ex, yshift=0.5mm, fill=white, draw=black, minimum height=15mm, yshift=-0.5mm, minimum width=5mm, font={\small}, tikzit category=circuit]

\tikzstyle{hadamard edge}=[-, dashed, dash pattern=on 2pt off 0.5pt, thick, draw={rgb,255: red,68; green,136; blue,255}]
\tikzstyle{box edge}=[-, dashed, dash pattern=on 2pt off 0.5pt, thick, draw={rgb,255: red,203; green,192; blue,225}]
\tikzstyle{brace edge}=[-, tikzit draw=blue, decorate, decoration={brace,amplitude=1mm,raise=-1mm}]
\tikzstyle{diredge}=[->]
\tikzstyle{double edge}=[-, double, shorten <=-1mm, shorten >=-1mm, double distance=2pt]
\tikzstyle{gray edge}=[-, {gray!60!white}]
\tikzstyle{pointer edge}=[->, very thick, gray]
\tikzstyle{boldedge}=[-, line width=1.6pt, shorten <=-0.17mm, shorten >=-0.17mm]
\tikzstyle{bidir edge}=[<->, very thick, draw={rgb,255: red,191; green,191; blue,191}]
\tikzstyle{separator edge}=[-, dashed, dash pattern=on 2pt off 0.5pt, thick, draw={rgb,255: red,153; green,153; blue,153}]
\tikzstyle{dashed edge}=[-, dashed, dash pattern=on 2pt off 0.5pt, thick]
\tikzstyle{solid edge}=[-, thick]

\DeclareMathOperator{\minpair}{minpair} 

\title{Smarter k-Partitioning of ZX-Diagrams for Improved Quantum Circuit Simulation}
\author{Matthew Sutcliffe
\institute{Department of Computer Science\\ University of Oxford\\ Oxford, UK}
\email{matthew.sutcliffe@cs.ox.ac.uk}
}
\def\titlerunning{Smarter k-Partitioning of ZX-Diagrams}
\def\authorrunning{M. Sutcliffe}
\begin{document}
\maketitle

\begin{abstract}
We introduce a novel method for strong classical simulation of quantum circuits based on optimally $k$-partitioning ZX-diagrams, reducing each part individually, and then efficiently cross-referencing their results to conclude the overall probability amplitude of the original circuit. We then analyse how this method fares against the alternatives for circuits of various size, shape, and interconnectedness and demonstrate how it is often liable to outperform those alternatives in speed by orders of magnitude.
\end{abstract}

\section{Introduction}

Quantum computers offer great promise for solving tasks in runtimes beyond the scope of their classical counterparts \cite{nielsen2010quantum}. But, as they exist today, their small scales and significant decoherence \cite{schlosshauer2019quantum} leave practical applications still out of reach. This leaves a gulf between proposed quantum software and available quantum hardware. In this space, classical simulation is very useful \cite{bravyi2016improved,BSS,jozsa2006simulation,huang2020classical,noh2020efficient}, allowing testing and verification of quantum algorithms and systems without the need for a quantum computer.

The drawback is that, without the quantum advantage, simulating such algorithms is necessarily very time-inefficient. Nevertheless, in recent years much research has been published on finding new techniques to improve this efficiency, enabling ever larger and more complex quantum circuits to be simulated. To this end, the graphical language of ZX-calculus \cite{coecke-duncan,van2020zx} (an alternative to the conventional quantum circuit notation \cite{nielsen2010quantum}) has proven to be very useful \cite{kissinger2022simulating1,kissinger2022simulating2,codsi2022classically,Sutcliffe2024,kochSim}. With a set of known rewriting rules and decompositions, any quantum circuit can be classically simulated by reducing its corresponding ZX-diagram.

In this paper, we introduce techniques by which ZX-diagrams can be reduced more efficiently via optimised graph partitioning. In particular, these methods allow a ZX-diagram to be broken down into smaller, more manageable parts which may be reduced far more efficiently. The reduced parts may then be iteratively regrouped pairwise until the whole is reformed fully reduced. In quantifying the results, we show how this approach can allow for a runtime reduction - to the task of classical simulation - of orders of magnitude and we explore how its effectiveness varies with the interconnectedness of the initial circuit.

\section{Background}

\subsection{ZX-calculus}

The ZX-calculus \cite{coecke-duncan,van2020zx} is a diagrammatic language for denoting quantum circuits as \textit{ZX-diagrams}, consisting of \textit{spiders} connected by \textit{edges}. Specifically, the spiders come in two varieties, namely green \textit{Z-spiders} and red \textit{X-spiders}, being linear operators defined as such:

\ctikzfig{spiderDefs}

Each spider has an associated \textit{phase}, $\alpha\in\mathbb{R}$ modulo $2\pi$, written either within or beside the spider itself (or typically neglected if zero). Each spider may have an arbitrary number of inputs and outputs (i.e. edges) and may be composed together to create a ZX-diagram. Moreover, these diagrams may be freely deformed and rearranged, provided the edge connectivity is preserved.

For convenience, the Hadamard gate is offered its own symbol (and an edge containing a Hadamard gate its own notation), though this too may be decomposed into Z- and X- spiders:

\ctikzfig{hadDef}

Allowing arbitrary spider phases, $[0,2\pi)$, enables ZX-diagrams to express any linear map, and indeed limiting spider phases to $n\pi/4$ (for $n\in\mathbb{Z}$) is sufficient for quantum completeness, meaning any quantum circuit may be expressed as a ZX-diagram with spiders of such phases.

\begin{figure}[bp!]
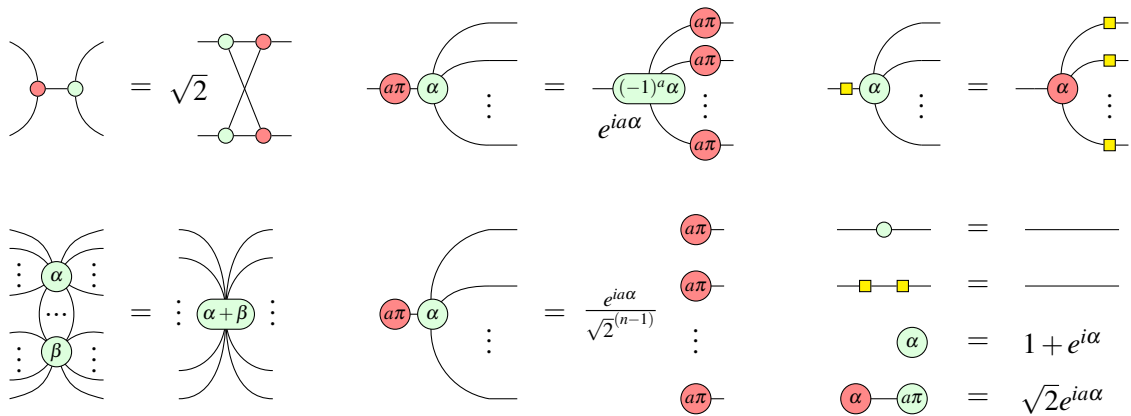

  \ctikzfig{zxrules}
  \caption{The complete \cite{jeandel2020completeness,backens2016completenesszxcalculus,wang2023completenesszxcalculus} set of fundamental rewriting rules \cite{van2020zx} in ZX-calculus, where $\alpha,\beta,\ldots\in\mathbb{R}$ and $a,b,\ldots\in\mathbb{B}$ (and $n$ is the number of output wires). These rules likewise hold with all colours inverted.}
  \label{fig:zxrules}
\end{figure}

The benefit of ZX-calculus over the traditional circuit notation is that it comes equipped with known \textit{rewriting rules} which describe equality relations between diagram patterns. The complete fundamental set of rewriting rules is shown in figure \ref{fig:zxrules}. Applying these rules where appropriate, a large and complex ZX-diagram may be simplified to an equivalent but smaller ZX-diagram (i.e. one containing fewer spiders). This is useful as, among other reasons, the simplified ZX-diagram may then be translated back into circuit notation, though now with fewer gates. In turn, reducing the number of gates helps minimise decoherence, and thus improve accuracy, when executed on quantum hardware \cite{nielsen2010quantum}.

\subsection{Classical simulation}

It is possible to simulate quantum circuits on classical hardware, though this comes at the cost of an exponential runtime. Nevertheless, it is a useful endeavour, used for - among other things - verifying the behaviour of quantum software and hardware, particularly in the present-day where the quantum computers available are often far too limited to execute such circuits. As such, much research has been published in recent years \cite{bravyi2016improved,BSS,jozsa2006simulation,huang2020classical,noh2020efficient} aimed at finding new techniques to help improve the efficiency of this task, in order to allow ever larger and more complex circuits to be simulated within feasible timeframes.

In particular, ZX-calculus has been employed to this end. By `plugging' all the inputs and outputs of a ZX-diagram with spiders, it becomes a \textit{scalar ZX-diagram}, which may be fully reduced (by the means outlined ahead) to a scalar expression. Specifically, one may plug the inputs and outputs with X-spiders corresponding to some input and output bitstrings (where spider phases of $0$ and $\pi$ correspond to bit inputs/outputs of $0$ and $1$ respectively). The resulting scalar, $A$, that this ZX-diagram is equal to relates to the probability, $P$, of measuring the specified output bitstring when executing the circuit given the specified input bitstring: $P=|A|^2$. This is known as `strong simulation', though can in turn be used, quite straightforwardly, to also perform `weak simulation' \cite{kissinger2022simulating1}, whereby one wishes to sample some output bitstring from a circuit based on its true probability distribution.

\textit{Clifford} ZX-diagrams - that is those with phases restricted to $n\pi/2$ (for $n\in\mathbb{Z}$) - can be simulated very efficiently. In fact, the rewriting rules outlined in figure \ref{fig:zxrules} are sufficient to fully reduce a scalar Clifford ZX-diagram to a scalar expression. However, this is not true for the computationally complete Clifford+T set, which includes \textit{T-spiders} (whose phases are of $n\pi/4$, for odd $n$). While the rewriting rules are still helpful in reducing the total number of spiders (including, often, many T-spiders), they ultimately reach a dead-end, resulting in a reduced ZX-diagram that may look like the following example:

\ctikzfig{exampleGraph}

To proceed from this stage, one may make use of \textit{stabiliser state decompositions} to exchange sets of T-spiders for some sum of \textit{stabiliser terms}. For instance, the following decomposition \cite{kissinger2022simulating2} allows a set of 2 T-spiders to be exchanged for a sum of 2 locally Clifford terms:

\ctikzfig{2to2decomp}

One can then repeat this process on the resulting terms to remove a further 2 T-spiders from each, and so on until one is left with a sum of fully Clifford terms which may each be reduced to scalars via the rewriting rules. Taking the sum of these scalars then provides the scalar result for the original Clifford+T circuit. The downside is this exponential growth in the number of terms to compute. In the case of the aforementioned decomposition, an initial (post Clifford simplification) ZX-diagram of $t$ T-spiders will translate to $2^{t/2}$ Clifford terms. Consequently, for a general $2^{\alpha t}$ terms, we say this decomposition has an efficiency of $\alpha=0.5$.

In practise, however, this is an upper-bound estimate of the true number of terms as one may apply inter-step simplification after each decomposition to help minimise the number of T-spiders (or `\textit{T-count}'). The present state of the art is the family of decompositions outlined in \cite{kissinger2022simulating2}, with a theoretical asymptotic efficiency of $\alpha\approx0.396$. (In practise, this family of decompositions, together with inter-step simplification, tends to achieve results closer to $\alpha\approx0.32$.)

\subsection{Graph partitioning}
\label{sec:background:partitioning}

It is possible, after initial Clifford simplification, to end up with a ZX-diagram consisting of two (or more) wholly separated subgraphs. Processing this na\"{i}vely as a single graph would mean decomposing it into $2^{\alpha t}$ terms. However, such graphs can be more efficiently processed by decomposing their partitioned subgraphs \textit{independently} and simply multiplying together their respective resulting scalars. The number of terms required is thus reduced to $2^{\alpha t_A} + 2^{\alpha t_B}$, where $t_A$ and $t_B$ are the T-counts of the respective subgraphs (hence $t=t_A+t_B$). In the best case, the partitioned segments would be of roughly equal T-counts, dramatically reducing the number of terms from $2^{\alpha t}$ to $2\cdot2^{\alpha t/2}$.

\begin{figure}
  \ctikzfig{Vcutting}
  \caption{The \textit{cutting} decomposition, which converts an arbitrary Z-spider into a sum of two Clifford X-spiders (or an arbitrary X-spider into a sum of two Clifford Z-spiders). The approximation symbol ($\approx$) denotes `equal up to a scalar', meaning there is a global scalar factor neglected here for brevity. Note that this decomposition, in conjunction with introducing a $0$-phase spider upon an edge (as is in the rules of figure \ref{fig:zxrules}), allows also for \textit{edges} to be cut into two terms.}
  \label{fig:cutting}
\end{figure}

In practise, however, such partitions rarely arise naturally. Nevertheless, one can evoke these partitions by cutting appropriate edges using the decomposition of figure \ref{fig:cutting}. This comes at the cost of a $2^c$ factor in the number of terms, given $c$ cuts \cite{Codsi2022Masters}. This is illustrated in figure \ref{fig:partition}. In fact, it is more efficient to cut \textit{vertices} rather than edges, as cutting a vertex can have the same effect as cutting all its edges individually \cite{Codsi2022Masters}. So the problem becomes trying to find a minimal set of vertices (i.e. spiders) to cut such that the ZX-diagram may be partitioned into $k$ independent parts of roughly equal T-counts

\begin{figure}[bp]
  \ctikzfig{partition01}
  \caption{A ZX-diagram may be partitioned into two or more disconnected parts (or \textit{segments}) at the cost of $2^c$ terms, given $c$ cuts.}
  \label{fig:partition}
\end{figure}


In solving this problem we may turn to graph theory literature. In particular, this is the \textit{vertex separator problem} \cite{rendl2018min,althoby2020exact}, but it is better mapped to the similar \textit{graph partitioning problem} \cite{cornaz2019vertex,andreev2004balanced,kuvcera1995expected,feder1999complexity}, which focuses on cutting edges rather than vertices, as the literature for this is more thoroughly established \cite{Codsi2022Masters}. As we wish to cut spiders rather than edges, we may translate our ZX-diagrams into \textit{hypergraphs} \cite{berge1984hypergraphs} by exchanging every edge for a vertex and every $n$-degree vertex for an $n$-ended hyperedge. We may then make use of existing implementations of state-of-the-art algorithms for efficiently partitioning a hypergraph $k$-ways. Specifically, we use the \textit{KaHyPar} \cite{DBLP:phd/dnb/Schlag20,ahss2017alenex} Python package, providing a weight of $1$ to each T-spider to balance T-count among the parts as best as possible. The specifics of how such partitioning algorithms work is beyond the scope of this paper, and for our purposes we may just consider KaHyPar's implementation an abstracted black box function for efficiently $k$-partitioning graphs. Nevertheless, we would direct interested readers to the relevant literature \cite{DBLP:phd/dnb/Schlag20,ahss2017alenex,schlag2015kwayhypergraphpartitioningnlevel,10.1145/3205455.3205475,henne2015nlevelhypergraphpartitioning}.


A major limitation in the existing literature of partitioning ZX-diagrams is that, in practise, partitioning into much more than 2 or 3 parts becomes too costly due to the large number of cuts required. Consequently, the literature is focused on bi-partitions (or at least $k$-partitions of very small $k$). The methods detailed ahead in this paper go some way to overcome this limitation.

\section{Methods}
\label{sec:methods}

The background discussed thus far represents the current extent of the literature on ZX-diagram partitioning, largely as put forward by \cite{Codsi2022Masters}. In the sections ahead, we further the state of the art by presenting three significant extensions of this work, ultimately culminating in drastic exponential speedups to the runtime of classical simulation for partitionable circuits.

\subsection{GPU-parallelised reduction}
\label{sec:methods:paramzx}

The first point we recognise is that after partitioning a graph via $c$ cuts, the resulting $2^c$ (as yet not decomposed to Clifford) ZX-diagram summand terms are identically structured, save for some Boolean-$\pi$ parameterised spiders. This immediately lends itself to the GPU-parallelised evaluation method outlined in the author's previous work \cite{sutcliffe2024fastclassicalsimulationquantum}.

To briefly summarise that work, we recognise that in these situations applying the standard Clifford simplification procedure to each of the graphs at this stage would break their structural commonality. So, instead we opt for a slightly modified, parameter-safe, version of the simplification routine which allows the graphs to maintain their common structure and to be decomposed to their respective scalars via efficient GPU-parallel computation. As shown in \cite{sutcliffe2024fastclassicalsimulationquantum}, this already can - in some cases - reduce the runtime by a factor of over $100$. And indeed, this linear speedup factor would be on top of the exponential speedup attained by the graph partitioning itself. One notable difference is that in the aforementioned paper this applies for many parallel instances of strong classical simulation, whereas here it would be applied to speeding up simulation of an individual circuit.

In short, by parameterising the cuts and maintaining consistent graph structure among all the summand terms, we need not compute them individually but rather can compute them in a parallel batches (limited only by the number of GPU cores available), to immediately gain a further notable speedup.


\subsection{Redundancy mitigation via parameterisation}

The next point we recognise is that the present means of classically simulating a ZX-diagram which has been partitioned into $k>2$ parts involves a lot of redundancy inherent among its calculations. We identify this redundancy and show how it can be negated to provide further drastic speedups in exchange for an increased memory overhead.

Consider, for instance, a ZX-diagram that has been partitioned into 4 disjoint parts via 9 cuts. In this example, there are 9 Boolean parameters (from the 9 cuts) to sum over, giving $2^{9}$ summand terms to compute. Graphically, with parameters $a,b,\ldots,i\in\{0,1\}$, this is:

\ctikzfig{regroupSeg_1}

This may be equivalently be expressed algebraically like so:

\begin{equation}
\sum_{a,b,\ldots,i}^{\{0,1\}}{ A(a,b,c) \cdot B(a,b,c,d,e,f) \cdot C(d,e,f,g,h,i) \cdot D(g,h,i) }
\label{eqn:abcdSummandTerms}
\end{equation}

Here, $A$, $B$, $C$, and $D$ are the independent partitioned subgraphs, each depending on some subset of the 9 parameters, $a,b,\ldots,i$. In the first term, all parameters are zero and hence the calculation is $A(000) \cdot B(000000) \cdot C(000000) \cdot D(000)$. Practically, this means substituting in zero for all of the parameters, then simplifying and decomposing each of the ZX-diagrams, $A(000)$, $B(000000)$, $C(000000)$, $D(000)$, before finally taking the product of the four resulting scalars.

The second summand term (where $a,\ldots,h,i = 0,\ldots,0,1$) would then be calculated likewise as $A(000) \cdot B(000000) \cdot C(000001) \cdot D(001)$. However, notice here that the parameters on which $A$ and $B$ depend have not changed, and so in calculating again $A(000)$ and $B(000000)$ we are simply repeating work that has already been done. And as each of these calculations involves fully decomposing the $A$ and $B$ diagrams into $2^{\alpha t_A}$ and $2^{\alpha t_B}$ Clifford terms to compute their final scalars, this work can be relatively slow.

What we propose instead is to initially precompute all the \textit{unique} states for each subraph and then their scalars can be recalled as needed without needing to recompute. In this example, this means precomputing the $2^3$ unique states of $A(a,b,c)$ by iterating over only the parameters local to $A$. Then we can likewise precompute the $2^6$ unique states of $B(a,b,c,d,e,f)$ by iterating over its 6 local parameters, and so on for $C$ and $D$. Then, when we calculate each summand of expression \ref{eqn:abcdSummandTerms}, we can simply call from memory the appropriate scalars to calculate their product, without needing to undergo further slow and redundant ZX-reduction.

With this approach, we avoid unnecessarily repeating slow ZX-calculus reductions by precomputing the unique terms by iterating over only the \textit{local} parameters, rather than all global parameters, for each graph. In our example, this brings the number of ZX-reductions from $2^9 = 512$ down to $2^3 + 2^6 + 2^6 + 2^3 = 144$, though this is merely a trivial case and in practise this is liable to offer many orders of magnitude reduction. However, despite minimising the slow part of the calculations, we still need to make $2^9$ calculations to cross-reference all these precomputed scalars. In other words, we still have $2^c$ summand terms, where $c$ is the global number of parameters (i.e. cuts) - we have just made each of these terms a lot less time-consuming to calculate. But, in the following subsection we show how even the number of cross-reference calculations (and hence the total number of calculations) can be exponentially reduced.

\subsection{Pairwise partition regrouping}
\label{sec:methods:regrouping}

This is most significant contribution we present as it results in not just a substantial linear reduction in the runtime, but an exponential one. This is the idea of pairwise regrouping of the partitioned segments. Returning to the example of the previous subsection, we noted that there are $2^9$ summand terms in this expression, with each being a product of four precomputed subterms (one for each partitioned segment).

Rather than computing the overall scalar directly, from summing over all parameters (as in expression \ref{eqn:abcdSummandTerms}), suppose instead one seeks initially to only regroup just two neighbouring segments. In the example case, this could mean regrouping segments $A$ and $B$ by summing over only the parameters common to them, while ignoring the other segments and parameters. Expressed algebraically, this is:

\[ \sum_{a,b,c}^{\{0,1\}}{ A(a,b,c) \cdot B(a,b,c,d,e,f) } = AB(d,e,f) \]

Recall that each segment is recorded as a list of scalars - one for each possible bitstring of its parameters. For example, $B(a,b,c,d,e,f)$ is recorded as a list of $2^6$ scalars. With this in mind, to ensure the result, $AB(d,e,f)$, is likewise recorded as a list of $2^3$ scalars, rather than some many-termed parameterised expression, it is important to sum over all the parameters \textit{local} to $A$ and $B$, and not just the ones common to both. As such, this step can be computed with $2^p$ calculations (in this case $2^6$), where $p$ is the number of \textit{local} parameters involved. (More on this shortly.)

In this example, we have $A(a,b,c)$ and $B(a,b,c,d,e,f)$, which collectively include 6 parameters. Thus, we can regroup these segments into $AB(d,e,f)$ via just $2^6$ calculations: 

\ctikzfig{regroupSeg_2}

By the same means, we can also regroup $C(d,e,f,g,h,i)$ and $D(g,h,i)$ into $CD(d,e,f)$ via $2^6$ calculations, giving us:

\ctikzfig{regroupSeg_3}

With one more iteration, we can now regroup $AB(d,e,f)$ and $CD(d,e,f)$ into $ABCD$ with just $2^3$ calculations:

\ctikzfig{regroupSeg_4}

Having now regrouped all segments into a single segment of no parameters, the result is a list containing just $2^0=1$ scalar. This is the scalar expression equivalent to the original ZX-diagram. Thus, we have reached the final answer via $2^6 + 2^6 + 2^3 = 136$ cross-reference calculations to regroup all the partitioned segments pairwise. This is opposed to $2^9 = 512$ cross-reference calculations as would have been required if we had just computed directly as per expression \ref{eqn:abcdSummandTerms}. Recall that this is a very simple illustrative example and that, in practise, the difference offered by this approach could be many orders of magnitude in the number of calculations.


As it is not strictly obvious or trivial, we will outline specifically what each such `cross-reference calculation' involves and how the segment pair regrouping procedure works. Consider, as the most trivial example, a chain of three partitioned segments, of which we want to regroup the first two:

\ctikzfig{regroupExample}

In this case, each segment contains just 2 parameters and so is recorded as a list of $2^2=4$ scalars. Of the 4 global parameters, $a,b,c,d$, three are local to segments $A$ and $B$ (namely $a,b,c$). It is these three parameters we will be summing over, for $2^3$ cross-reference calculations - i.e. $a,b,c=0,0,0$, then $a,b,c=0,0,1$, and so on. In each case, we may retrieve the scalars $A(a,b)$ and $B(b,c)$ and multiply them to deduce the scalar $AB(a,b,c)$. In turn, this newly calculated scalar may be added to the relevant $AB(a,c)$. That is to say, after computing all products, $AB(a,b,c) \forall (a,b,c)$, we can reduce the resulting list to $AB(a,c)$ by summing $\sum_{b}^{\{0,1\}}{AB(a,b,c)} = AB(a,c)$ for each $(a,c)$. This is illustrated in table \ref{table:regroup}, and simple high-level pseudocode that implements this procedure is shown in algorithm \ref{alg:regroupPseudo}. An alternative highly efficient low-level and GPU-parallelised implementation is shown in appendix \ref{app:cudacode}.

\begin{table}[]
\centering
\begin{tabular}{|c
>{\columncolor[HTML]{C0C0C0}}c c|cc|c|cc|}
\cellcolor[HTML]{000000}{\color[HTML]{FFFFFF} \textbf{$a$}} & \cellcolor[HTML]{000000}{\color[HTML]{FFFFFF} \textbf{$b$}} & \cellcolor[HTML]{000000}{\color[HTML]{FFFFFF} \textbf{$c$}} & \cellcolor[HTML]{000000}{\color[HTML]{FFFFFF} \textbf{$A_{ab}$}} & \cellcolor[HTML]{000000}{\color[HTML]{FFFFFF} \textbf{$B_{bc}$}} & \cellcolor[HTML]{000000}{\color[HTML]{FFFFFF} \textbf{$(AB)_{abc}$}} & \cellcolor[HTML]{000000}{\color[HTML]{FFFFFF} \textbf{$\sum\limits_{b}^{\{0,1\}}{(AB)_{abc}}$}} & \cellcolor[HTML]{000000}{\color[HTML]{FFFFFF} \textbf{$\equiv(AB)_{ac}$}} \\ \hline
\cellcolor[HTML]{FFCE93}{\color[HTML]{333333} $0$}          & {\color[HTML]{333333} $0$}                                  & \cellcolor[HTML]{FFCE93}{\color[HTML]{333333} $0$}          & {\color[HTML]{333333} $A_{00}$}                                  & {\color[HTML]{333333} $B_{00}$}                                  & \cellcolor[HTML]{FFCE93}{\color[HTML]{333333} $A_{00}B_{00}$}        & \cellcolor[HTML]{FFCE93}{\color[HTML]{333333} $(A_{00}B_{00}+A_{01}B_{10})$}                  & \cellcolor[HTML]{FFCE93}$\equiv(AB)_{00}$                                 \\
\cellcolor[HTML]{FFFFC7}{\color[HTML]{333333} $0$}          & {\color[HTML]{333333} $0$}                                  & \cellcolor[HTML]{FFFFC7}{\color[HTML]{333333} $1$}          & {\color[HTML]{333333} $A_{00}$}                                  & {\color[HTML]{333333} $B_{01}$}                                  & \cellcolor[HTML]{FFFFC7}{\color[HTML]{333333} $A_{00}B_{01}$}        & \cellcolor[HTML]{FFFFC7}{\color[HTML]{333333} $(A_{00}B_{01}+A_{01}B_{11})$}                  & \cellcolor[HTML]{FFFFC7}$\equiv(AB)_{01}$                                 \\
\cellcolor[HTML]{FFCE93}{\color[HTML]{333333} $0$}          & {\color[HTML]{333333} $1$}                                  & \cellcolor[HTML]{FFCE93}{\color[HTML]{333333} $0$}          & {\color[HTML]{333333} $A_{01}$}                                  & {\color[HTML]{333333} $B_{10}$}                                  & \cellcolor[HTML]{FFCE93}{\color[HTML]{333333} $A_{01}B_{10}$}        & \cellcolor[HTML]{C0C0C0}{\color[HTML]{333333} }                                               & \cellcolor[HTML]{C0C0C0}                                                  \\
\cellcolor[HTML]{FFFFC7}{\color[HTML]{333333} $0$}          & {\color[HTML]{333333} $1$}                                  & \cellcolor[HTML]{FFFFC7}{\color[HTML]{333333} $1$}          & {\color[HTML]{333333} $A_{01}$}                                  & {\color[HTML]{333333} $B_{11}$}                                  & \cellcolor[HTML]{FFFFC7}{\color[HTML]{333333} $A_{01}B_{11}$}        & \cellcolor[HTML]{C0C0C0}{\color[HTML]{333333} }                                               & \cellcolor[HTML]{C0C0C0}                                                  \\
\cellcolor[HTML]{9AFF99}{\color[HTML]{333333} $1$}          & {\color[HTML]{333333} $0$}                                  & \cellcolor[HTML]{9AFF99}{\color[HTML]{333333} $0$}          & {\color[HTML]{333333} $A_{10}$}                                  & {\color[HTML]{333333} $B_{00}$}                                  & \cellcolor[HTML]{9AFF99}{\color[HTML]{333333} $A_{10}B_{00}$}        & \cellcolor[HTML]{9AFF99}{\color[HTML]{333333} $(A_{10}B_{00}+A_{11}B_{10})$}                  & \cellcolor[HTML]{9AFF99}$\equiv(AB)_{10}$                                 \\
\cellcolor[HTML]{CBCEFB}{\color[HTML]{333333} $1$}          & {\color[HTML]{333333} $0$}                                  & \cellcolor[HTML]{CBCEFB}{\color[HTML]{333333} $1$}          & {\color[HTML]{333333} $A_{10}$}                                  & {\color[HTML]{333333} $B_{01}$}                                  & \cellcolor[HTML]{CBCEFB}{\color[HTML]{333333} $A_{10}B_{01}$}        & \cellcolor[HTML]{CBCEFB}{\color[HTML]{333333} $(A_{10}B_{01}+A_{11}B_{11})$}                  & \cellcolor[HTML]{CBCEFB}$\equiv(AB)_{11}$                                 \\
\cellcolor[HTML]{9AFF99}{\color[HTML]{333333} $1$}          & {\color[HTML]{333333} $1$}                                  & \cellcolor[HTML]{9AFF99}{\color[HTML]{333333} $0$}          & {\color[HTML]{333333} $A_{11}$}                                  & {\color[HTML]{333333} $B_{10}$}                                  & \cellcolor[HTML]{9AFF99}{\color[HTML]{333333} $A_{11}B_{10}$}        & \cellcolor[HTML]{C0C0C0}{\color[HTML]{333333} }                                               & \cellcolor[HTML]{C0C0C0}                                                  \\
\cellcolor[HTML]{CBCEFB}{\color[HTML]{333333} $1$}          & {\color[HTML]{333333} $1$}                                  & \cellcolor[HTML]{CBCEFB}{\color[HTML]{333333} $1$}          & {\color[HTML]{333333} $A_{11}$}                                  & {\color[HTML]{333333} $B_{11}$}                                  & \cellcolor[HTML]{CBCEFB}{\color[HTML]{333333} $A_{11}B_{11}$}        & \cellcolor[HTML]{C0C0C0}{\color[HTML]{333333} }                                               & \cellcolor[HTML]{C0C0C0}{\color[HTML]{C0C0C0} peekaboo}                                                 \\ \hline
\end{tabular}
\caption{The lists of scalars representing two segments, $A_{ab}$ and $B_{bc}$, may be regrouped into $(AB)_{ac}$ by iterating over the parameters involved and cross-referencing the scalars accordingly.}
\label{table:regroup}
\end{table}

\begin{algorithm}[b!]
\caption{A high-level implementation of the pairwise segment regrouping function}
\begin{algorithmic}

\Function{regroup_pair}{$segA,segB$}
    \State $commonParams \gets segA.localParams \;\cup\; segB.localParams$
    \State $exclusiveParams \gets segA.localParams \;\Delta^*\; segB.localParams$
    \Comment{\textcolor{gray}{$//$ *The list of parameters that are common to $segA$ and $segB$ but are \textit{not} in both unless they are also in another segment: $(A \Delta B) \cup (A \cap B \cap (C \cup D \cup \ldots))$ $//$}}
    
    \State $n \gets \texttt{length}(commonParams)$
    \State $m \gets \texttt{length}(exclusiveParams)$
    \State $newScalars = [(0+0i)]\times2^{m}$
    
    \For{$i \gets 0$ to $2^{n}-1$}
        \State $bitstr \gets \texttt{dec_to_bin}(i,n)$
        \Comment{\textcolor{gray}{$//$ Convert decimal, $i$, to $n$-bit binary string $//$}}
        
        \State $ab \gets \texttt{localise_bits}(bitstr,commonParams,segA.localParams)$
        \Comment{\textcolor{gray}{$//$ $\texttt{localise_bits}(b,P,p)$ takes some bitstring, $b$, with bits corresponding to some set of parameters, $P$, and returns the bitstring which only includes the parameters in the subset $p$, e.g. $\texttt{localise_bits}(``110",\;[A,B,C],\;[A,C]) \rightarrow ``10"$ $//$}}
        \State $A_{ab} \gets segA.scalars[\texttt{bin_to_dec}(ab)]$
        
        \State $bc \gets \texttt{localise_bits}(bitstr,commonParams,segB.localParams)$
        \State $B_{bc} \gets segB.scalars[\texttt{bin_to_dec}(bc)]$

        \State $AB_{abc} \gets A_{ab} \times B_{bc}$
        \State $ac \gets \texttt{localise_bits}(bitstr,commonParams,exclusiveParams)$
        \State $newScalars[\texttt{bin_to_dec}(ac)] \gets newScalars[\texttt{bin_to_dec}(ac)] + AB_{abc}$
    \EndFor

    \State $segAB \gets \texttt{new} \;\: Segment()$
    \State $segAB.localParams \gets exclusiveParams$
    \State $segAB.scalars \gets newScalars$
    
    \State $segA \gets segAB$
    \State $segB \gets \texttt{null}$
\EndFunction

\end{algorithmic}
\label{alg:regroupPseudo}
\end{algorithm}


The simple examples discussed so far have been of neatly partitioned chains of segments. Realistically, however, efficiently partitioning a graph k-ways is likely to result in more chaotic and intertwined segment connections. In this context, a connection between (or among) segments represents a cut that has been made which gave rise to a parameter common to these segments. Moreover, as it is vertices (i.e. spiders) rather than edges that are being cut, it is possible for a cut to affect (and hence introduce a new parameter to) more than two segments. For relative neatness, we notate partitioned segments, and the connections among them, as a hypergraph like so:

\ctikzfig{connectivityFormat}

This should not be confused with the hypergraph representation of a ZX-diagram as discussed in section \ref{sec:background:partitioning}.

Given the goal is to eventually regroup all segments together, to do this efficiently (that is, minimising the number of calculations involved) we, at each step, regroup the pair which collectively have the fewest number of local parameters (i.e. hyperedges). For a hypergraph, $H$, containing $k$ nodes (i.e. segments), this number will be given by the function $\minpair{(H)}$, which can be computed in $O(k^2)$ time (and as $k$ is always relatively low, this runtime is generally negligible).


Given the methods presented in this paper, the number of computations required to fully reduce a $t$ T-gate ZX-diagram to its scalar has been brought down from $2^{\alpha t}$ to a potentially much more modest $S_{precomp} + S_{crossref}$ (plus some negligible overhead from the partitioning function itself), where:

\[ S_{precomp} \;=\; \sum_{i=1}^{k}{2^{\alpha t_i + c_i}} \]
\[ S_{crossref} \;=\; \sum_{i=0}^{k-2}{2^{\minpair{(H_i)}}} \]

The first equation here describes the computational cost of precomputing the unique scalars of each partitioned segment. $i$ iterates through each segment, such that $t_i$ and $c_i$ are its local T-count and local parameter (i.e. hyperedge) count respectively. Meanwhile, the second equation describes the computational cost of cross-referencing these precomputed scalars (as in table \ref{table:regroup}). In other words, it is the cost of regrouping the partitioned segments. Here, $i$ denotes the regrouping \textit{step} - that is, $i=0$ refers to the initial hypergraph state (with $k$ segments) and each successive step ($i\rightarrow i+1$) is defined by when the next cheapest pair of segments is regrouped (reducing the number of segments by one: $k\rightarrow k-1$). $H_i$, therefore, is the state of the hypergraph after $i$ instances of pairwise regrouping.

\subsection{The ZX-Partitioner}

All this is brought together in a Python package we developed called the \textit{ZX-Partitioner}, which may be found at: \href{https://github.com/mjsutcliffe99/zxpartitioner}{https://github.com/mjsutcliffe99/zxpartitioner}. At its most abstracted, this package offers a function into which the user may provide a ZX-diagram (a \textit{graph} from the \textit{PyZX} package \cite{pyzx}) and its scalar equivalent will be calculated using the methods outlined in this paper. For convenience and to help the reader/user better understand the methodology, this routine can also be run one phase at a time and with visualisations of the partitioned ZX-diagrams and, more helpfully, their segment connectivity hypergraph at each step. It may also be easily configured for different stabiliser state decomposition strategies and hardware capabilities.


Of this main function, the initial step performed is to determine the most efficient number of parts, $k$, into which the ZX-diagram should be partitioned. Partitioning into a greater number of parts means that each will be of a lower T-count and hence the number of \textit{precomputing} calculations, $S_{precomp}$, will be drastically reduced. (For a given graph, the typical number of \textit{local} cuts, $c_i$, on each part, $i$, tends to not vary too drastically regardless of $k$, and at any rate it is likely to be much smaller than $\alpha t_i$, so the local T-count tends to be the significant contributor to $S_{precomp}$. This is especially true when we consider the projected \textit{runtimes} rather than the number of computations, as each computation in the $2^{c_i}$ component can be computed much more rapidly than those in the $2^{\alpha t_i}$ component, as highlighted in section \ref{sec:methods:paramzx}.) However, taking a larger $k$ comes at the cost of increasing the number of \textit{total} cuts, $C$. While ordinarily this would render larger $k$ values infeasible, as we showed in section \ref{sec:methods:regrouping} this need not be the case. Nevertheless, taking a larger $k$ indeed increases the number of cross-reference calculations, $S_{crossref}$, albeit not so drastically. This is because when a pair of segments, $A$ and $B$, is regrouped, the resulting segment will have a number of local cuts, $c_{AB}$, equal to the number of cuts in the symmetric difference of $c_A$ and $c_B$. This in turn means that as more segments are regrouped pairwise, there is a higher likelihood of segments having larger numbers of local cuts, which results in a larger $\minpair{(H)}$.


Put concisely, partitioning into a larger $k$ results in decreasing $S_{precomp}$ but increasing $S_{crossref}$. As the \textit{overall} number of computations is given by the sum of these two, then the most optimal $k$ is that which produces the crossover point where these two terms are as close to equal as possible, such that neither dominates and renders the other negligible. (Note that the $k$-partitioning function itself generally runs in negligible time.) Fortunately, for any $k$, $S_{precomp}$ and $S_{crossref}$ can be determined in advance in negligible time. Consequently, the optimal choice of $k$ can likewise be determined in advance in negligible time. (Balancing projected \textit{runtimes}, $T_{precomp}$ and $T_{crossref}$, as in appendix \ref{app:projectedRuntimes}, yields even better results.)

\begin{figure}
  \centering
  \includegraphics[scale=0.5]{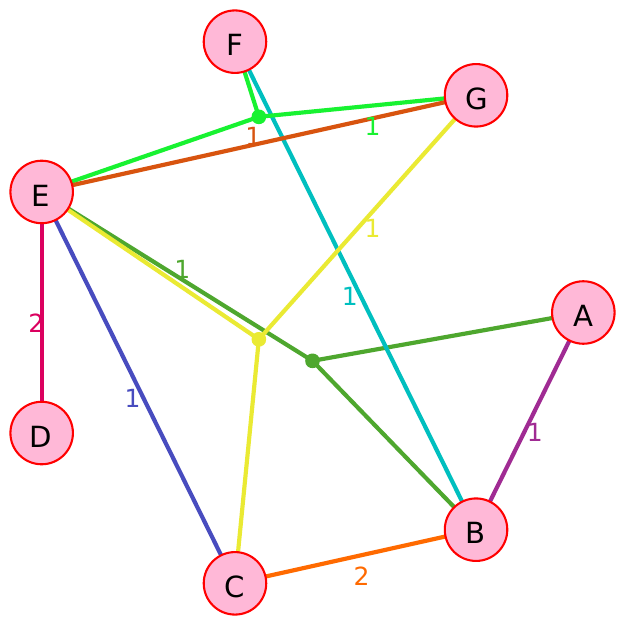}
  \caption{An example of a segment connectivity hypergraph, generated via the \textit{ZX-Partitioner}.}
  \label{fig:hNet0}
\end{figure}

With the optimal $k$ determined, the next step is to $k$-partition the ZX-diagram, which we can visualise as a hypergraph of partitioned segments with connected edges representing common cut parameters, as in figure \ref{fig:hNet0}.

At this point, the segments may be precomputed - in each case, $i$, turning a parameterised ZX-diagram (of $c_i$ parameters, arising from $c_i$ local cuts) into a list of $2^{c_i}$ scalars. $c_i$ here also denotes the number of edges connected to the particular segment, $i$, in the hypergraph (figure \ref{fig:hNet0}).

\begin{figure}[t]
  \centering
  \includegraphics[scale=0.45]{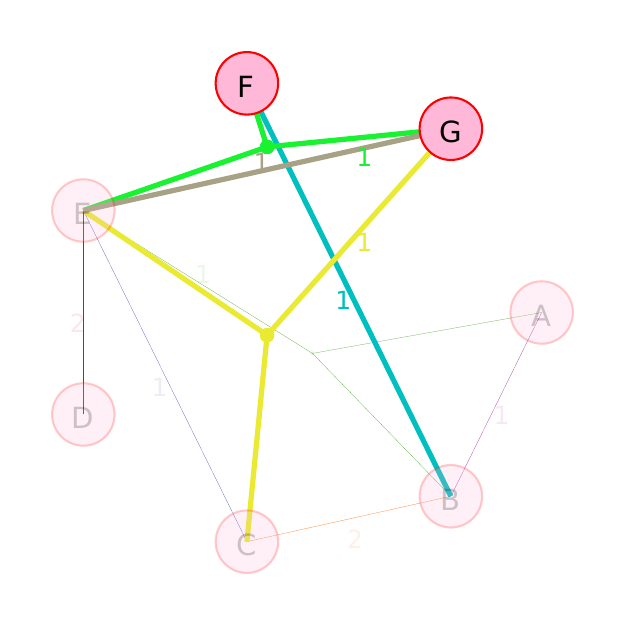}
  \includegraphics[scale=0.45]{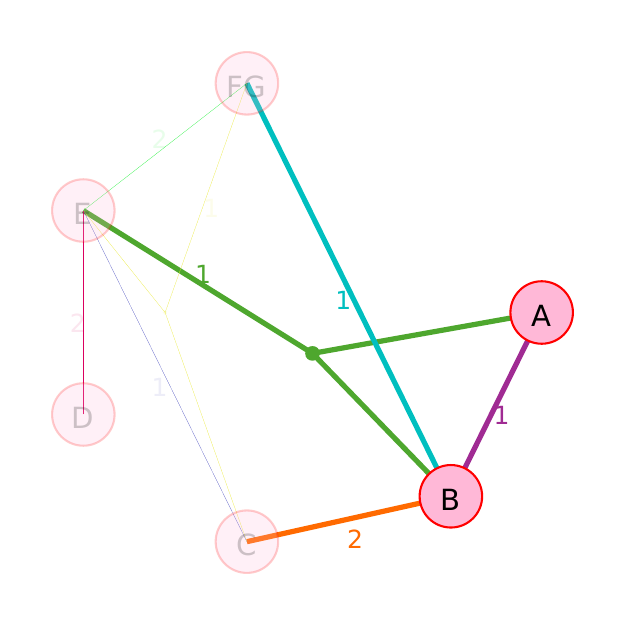}
  \includegraphics[scale=0.45]{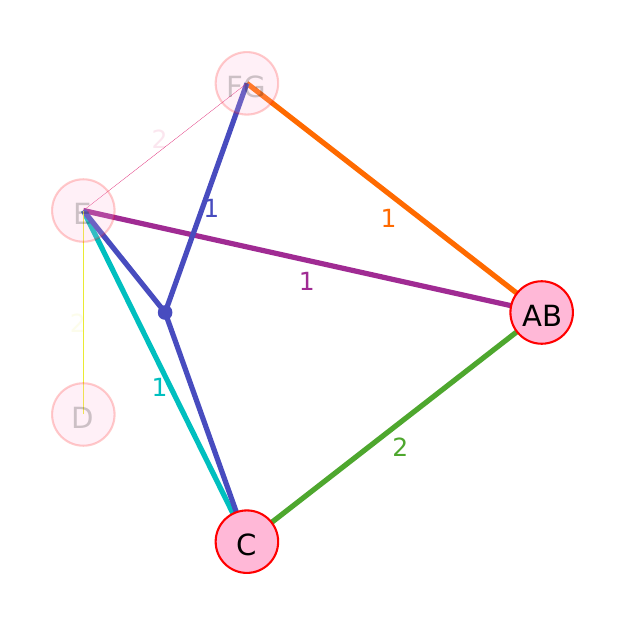}
  \\
  \includegraphics[scale=0.45]{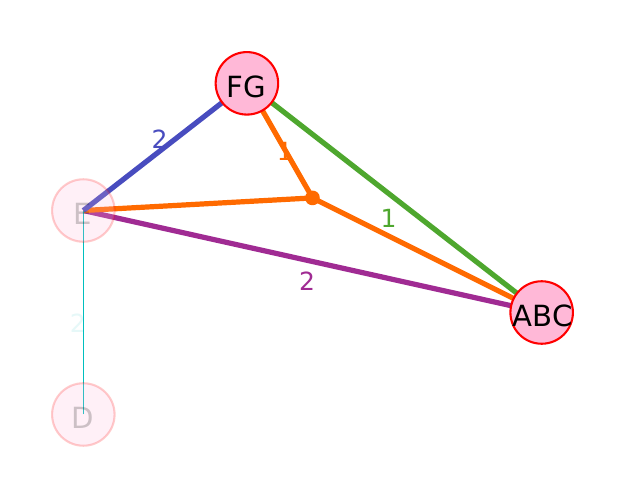}
  \includegraphics[scale=0.45]{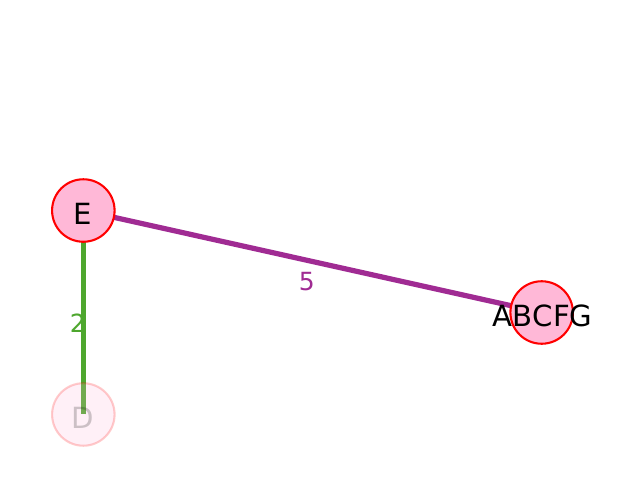}
  \includegraphics[scale=0.45]{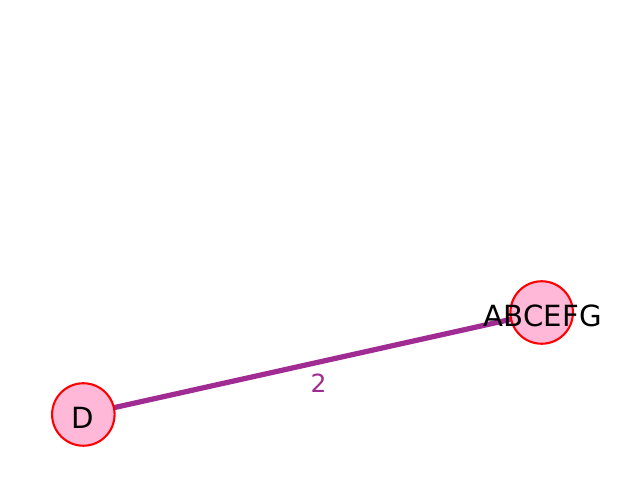}
  \caption{The precomputed segments of a partitioned ZX-diagram may be regrouped pairwise (selecting the cheapest pair to regroup at each step) until one segment remains. The steps in this figure are shown chronologically in row-major order. In each case, the local edges among the cheapest pair are highlighted, with the sum of their weights, $w$, giving the computational cost of regrouping, $2^w$. Regrouping the final remaining pair will provide the overall scalar result. (Note that the edge colours are random and exist for visual clarity but bear no meaning.)}
  \label{fig:hnetReduce}
\end{figure}

Next, the program will find the pair of connected segments, $A$ and $B$, with the fewest collective number of local edges, $c_A+c_B$. This will be the cheapest connected pair to regroup and so regrouped it is, into segment $AB$ (as detailed in section \ref{sec:methods:regrouping}). Having fused these two segments together, the hypergraph will now contain one fewer segment in total. This step may then be repeated, regrouping whichever connected pair of segments is \textit{now} the cheapest. This process continues until the final two segments are regrouped into one. Figure \ref{fig:hnetReduce} shows this in action.

This final segment will have no edges (i.e. local cuts) and hence will record a single (as $2^0 = 1$) scalar. This scalar is the final result, which is equivalent to the original scalar ZX-diagram.


\section{Results}

To benchmark the effectiveness of the ZX-Partitioner, we compared its projected runtime (see appendix \ref{app:projectedRuntimes}) for fully reducing random Clifford+T diagrams to that of directly reducing them via stabiliser decompositions \cite{kissinger2022simulating2} (with no partitioning). We also benchmarked the same random dataset for a \textit{na\"{i}ve} partitioning method which $k$-partitions (for its own self-determined optimal $k$) but does not apply the techniques outlined in section \ref{sec:methods}. In particular, we constructed many $n$-qubit circuits (of various $n$) by randomly placing gates of the set $\{T,S,HSH,CNOT\}$ with equal probability, up to the count of $d$ gates (which we call its \textit{depth}). (This is the \textit{generate.cliffordT} function of PyZX.) The circuits were then plugged with $\bra{+}^{\otimes n}$ and $\ket{+}^{\otimes n}$ to turn them into scalar diagrams, before finally they underwent an initial round of Clifford simplification. The resulting ZX-diagram, in each case, was taken as its initial state for the benchmarks, with the goal of each method being to fully reduce it to a scalar.

\begin{figure*}[t!]
    \centering
    \begin{subfigure}[t]{0.5\textwidth}
        \centering
        \includegraphics[scale=0.6]{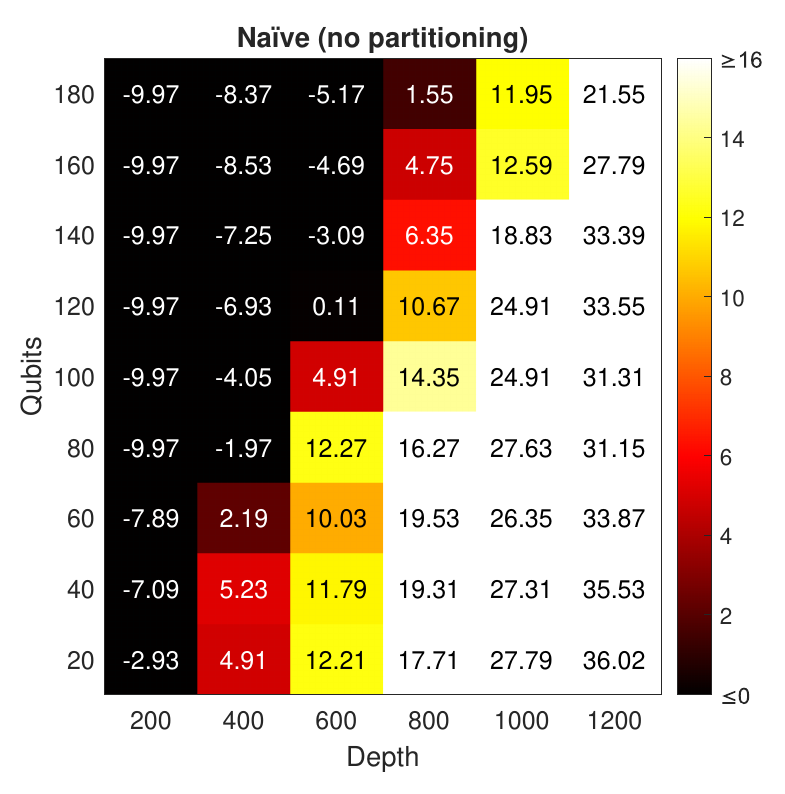}
    \end{subfigure}%
    ~ 
    \begin{subfigure}[t]{0.5\textwidth}
        \centering
        \includegraphics[scale=0.6]{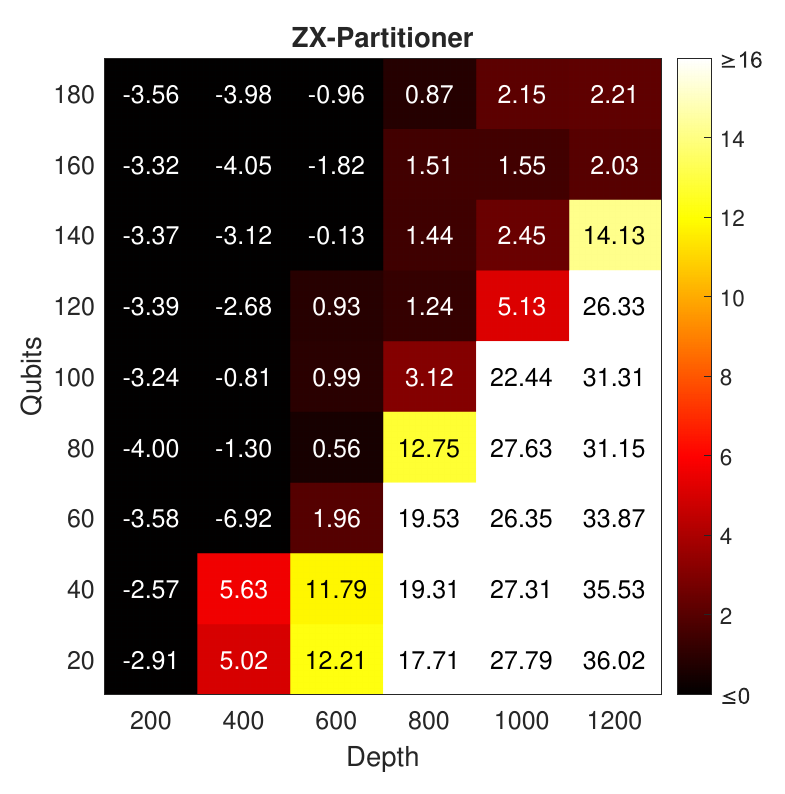}
    \end{subfigure}
    \caption{The average $log_2$ runtimes (in seconds) for fully reducing a ZX-diagram via (left) the direct decomposition approach and (right) the smart partitioning approach, for uniformly random Clifford+T circuits of various depths and qubit counts.}
    \label{fig:heatInf}
\end{figure*}

Figure \ref{fig:heatInf} shows the $log_2$ of the projected runtime results (in seconds) for each method on the random dataset of diagrams, varying in depth and number of qubits. The scale is $log_2$ such that, approximately speaking, $0$ represents a second, $6$ a minute, and $12$ an hour. $16$, meanwhile, represents just over $18$ hours, which we take as the rough upper limit of what is computationally feasible. Thus, in these heatmaps, black denotes trivial cases while white denotes practically incomputable cases. The coloured region in between then represents the region of interest, which we may call the \textit{frontier}.

Note that for each circuit size, we generate and compute $10$ samples and take the average result. The first thing one might notice is that the ZX-Partitioner never performs meaningfully slower than the direct decomposition approach. This is because the latter can be seen as a special case of the former, whereby the most optimal $k=1$ (that is, where any amount of partitioning would result in worse performance and so the reduction proceeds without any). More significantly, the figure shows that, for certain sizes of circuits, the ZX-Partitioner method outperforms the na\"{i}ve approach by many orders of magnitude.

Evidently, the ZX-Partitioner is most effective for many-qubit circuits of low depth, as well as few-qubit circuits of any depth. In fact, in the latter case, if initial Clifford simplification were avoided and partitions were enforced along qubit-lines, then it would always be possible to partition an $n$-qubit circuit into arbitrarily many parts, connected linearly in a chain, where each part connects to the next via $n$ edges. Consequently, with just slight modification, the smart partitioning approach could always achieve a runtime proportional to $O(2^{2n})$. However, few-qubit circuits are already known to be efficiently simulable (by computing the state vector \cite{jones2019quest,wu2019full} or via tensor contraction \cite{markov2008simulating,brennan2021tensor}). This leaves the more interesting case of shallow many-qubit circuits. It is easy to understand why these circuits are also particularly effective for the smart partitioner approach, as `depth' in this context refers to the total number of gates. Hence, when the ratio of the depth to the number of qubits is low, this describes circuits with few gates per qubit, and hence few CNOTs connecting these qubits, meaning few cuts would likely be needed to partition along these lines.

Beyond these cases, the ZX-Partitioner appears to offer no improvement versus direct decomposition. However, recall that this dataset was generated completely randomly, and so it is understandable that the frequency of good vertex cuts for partitioning shrank as the overall size of the graphs grew. In more realistic circuits, one would expect more inherent structure and, as such, a less sporadic placement of CNOTs. To try to model this with a new randomly generated dataset, we make a slight modification in how we place place the CNOTs. Whereas previously both ends of the CNOT were placed on different random qubits, we now instead place one end of the CNOT on a random qubit and then randomly decide the qubit of its other end according to a non-uniform probability distribution, to favour nearer qubits over further ones. Specifically, when deciding where to place the target qubit of the CNOT, we weigh the probabilities of the qubits according to a normal distribution about the control qubit, such that the probability of a CNOT spanning $\Delta q\geq1$ qubits is given by:

\[ P(\Delta q) = \frac{1}{\sqrt{2\pi\sigma^2}} e^{-\frac{(\Delta q-1)^2}{2\sigma^2}} \]

\begin{figure*}[t!]
    \centering
    \begin{subfigure}[t]{0.48\textwidth}
        \centering
        \includegraphics[scale=0.54]{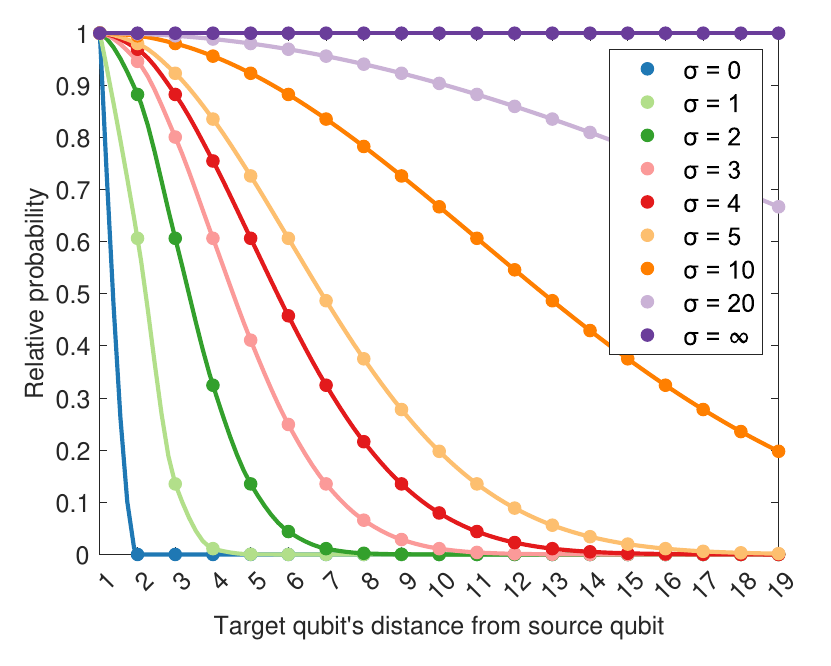}
        \caption{The probability amplitudes for a CNOT spreading $\Delta q$ qubits, according to a normal distribution with a standard deviation, $\sigma$.}
        \label{fig:normDist}
    \end{subfigure}\hfill%
    ~ 
    \begin{subfigure}[t]{0.48\textwidth}
        \centering
        \includegraphics[scale=0.54]{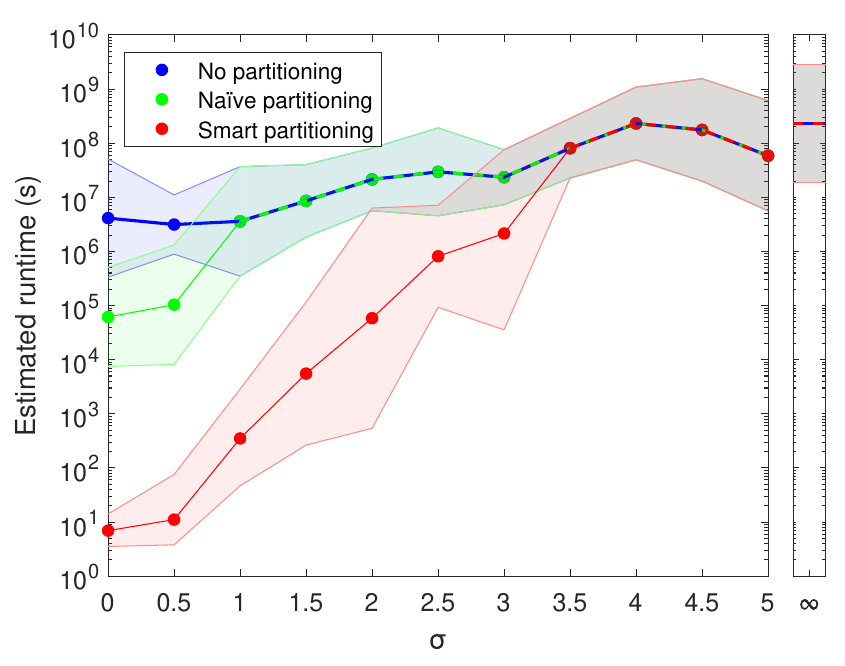}
        \caption{The average projected runtimes for strongly simulating Clifford+T circuits of $30$ qubits and a depth (gate count) of $1,000$, using three different methods.}
        \label{fig:timevsigmaCapped}
    \end{subfigure}
    \caption{We generated random Clifford+T circuits whereby the spread of each CNOT was decided probabilistically according to a normal distribution. By adjusting the variance of this distribution (left), the runtime for strongly classically simulating the circuits (right) is also affected.}
\end{figure*}

This is derived from the general form of the normal distribution function, where $\sigma$ denotes the standard deviation (and hence $\sigma^2$ denotes the variance). Figure \ref{fig:normDist} shows what this distribution looks like for various values of $\sigma$. Note that these have each been scaled (un-normalised) to show the probabilities relative to that of $P(\Delta q = 1)$. This way they are all clearly readable within the same plot. For instance, when $\sigma=3$ the span of a CNOT (i.e. the distance between its control and target) is roughly $0.6$ times as likely to be exactly $4$ qubits as it is to be exactly $1$ qubit. Moreover, when $\sigma=0$, this means that CNOTs always connect to their nearest neighbouring qubit (either immediately above or immediately below, with equal probability). Meanwhile, at the other extreme, when $\sigma=\infty$, the target of each CNOT will be placed on any of the qubits in the circuit with uniformly equal probability (as in the experiments of figure \ref{fig:heatInf}).

Given this modification, we consider again - as an illustrative example - randomly generated circuits of $30$ qubits and depth $1,000$. We repeat this experiment for various values of $\sigma$, in each case taking the averaged $log$ runtime over $10$ repeats. We likewise test the direct decomposition (i.e. no partitioning) method and the na\"{i}ve partitioning method against the same dataset. The results are shown in figure \ref{fig:timevsigmaCapped}, with error bars given by standard deviation of the $log$ runtimes over the $10$ repeats for each $\sigma$.

From this figure one will immediately observe that the effectiveness of the smart partitioning method is heavily impacted by $\sigma$. For circuits of this particular size, we notice that the method generally outperforms both the `no partitioning' and `na\"{i}ve partitioning' approaches (often by many orders of magnitude) when $\sigma\lessapprox3$. Indeed, as lower values of $\sigma$ are used to generate the random circuits, this improvement becomes ever more drastic, though even for very small $\sigma$, the smart partitioner's runtime doesn't fall much below a second, despite what would be predicted by estimating the runtimes from the number of precomputing and cross-referencing calculations. This is because, in these very speedy cases, the overhead runtime from the partitioning function itself (which can take up to a few seconds in the most extreme cases) is no longer negligible.

Moreover, the direct decomposition (`no partitioning') approach is relatively consistent as $\sigma$ is varied (as compared to the partitioning methods). This is because the complexity of this approach depends primarily upon the initial T-count, which itself is influenced by the distribution of CNOTs in only very roundabout ways. Additionally, neither partitioning method ever performs meaningfully slower than the na\"{i}ve approach because, as noted earlier, `no partitioning' is essentially the $k=1$ special case of partitioning (though, for interest, figure \ref{fig:timevsigmaUncapped} in appendix \ref{app:additionalResults} shows the equivalent results if the partitioning methods are forced to make at least one partition, such that $k>1$).

Furthermore, we see here that once the partitioning method has been capped by the na\"{i}ve approach (that is, when no partitioning becomes optimal) it tends to remain so as $\sigma$ is further increased. Lastly, as $\sigma\rightarrow\infty$ the methods each reach their terminal runtimes. In this case - because $k=1$ is deemed optimal - they each share the same terminal runtime, namely $10^{8.36\pm1.09}$ seconds (which is consistent with what was observed in figure \ref{fig:heatInf}). However, as evidenced by figure \ref{fig:heatInf}, there are circuit sizes for which even $\sigma=\infty$ leads to the smart partitioner method reigning supreme, and in such cases - if plotted against $\sigma$ like in figure \ref{fig:timevsigmaCapped} - one would observe that the smart partitioning method always remains below the others. One such example is shown in figure \ref{fig:timevsigmaB} in appendix \ref{app:additionalResults}.

\begin{figure*}[t!]
    \centering
    \begin{subfigure}[t]{0.5\textwidth}
        \centering
        \includegraphics[scale=0.6]{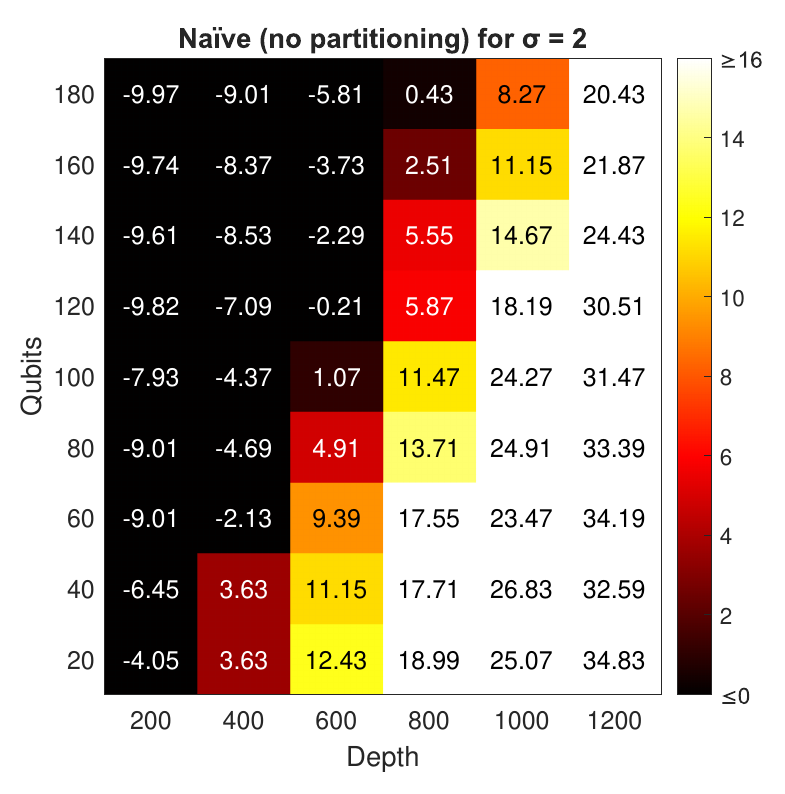}
    \end{subfigure}%
    ~ 
    \begin{subfigure}[t]{0.5\textwidth}
        \centering
        \includegraphics[scale=0.6]{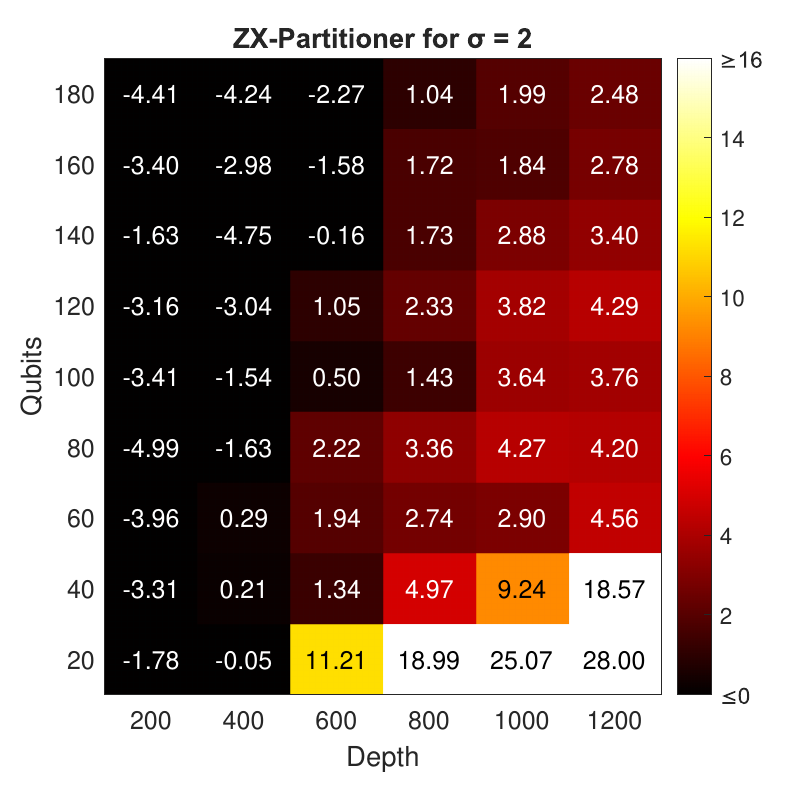}
    \end{subfigure}
    \caption{The average $log_2$ runtimes (in seconds) for fully reducing a ZX-diagram via (left) the direct decomposition approach and (right) the smart partitioning approach, for random Clifford+T circuits of various depths and qubit counts, with the spread of each CNOT given by a normal distribution with $\sigma=2$.}
    \label{fig:heatSig2}
\end{figure*}

Lastly, we repeated the figure \ref{fig:heatInf} experiments but for randomly generated circuits with $\sigma=2$ (as opposed to $\sigma=\infty$). These results are shown in figure \ref{fig:heatSig2}. Under these conditions, the na\"{i}ve approach has scarcely changed, yet the ZX-Partitioner shows significantly reduced runtimes, with a far shallower frontier. Clearly, therefore, CNOTs with a more localised spread (i.e. a lower $\sigma$) lead to circuits which are much more partitionable and hence even more suitable for such methods. Indeed, given low $\sigma$, we observe that for all bar the most unfavourably sized circuits, the partitioner method offers orders of magnitudes reduction to the runtime versus the na\"{i}ve alternative. (Furthermore, in appendix \ref{app:compoundCircuits} we discuss how our results compare to those achieved by tensor contraction and consider a slightly modified type of randomly generated circuit which is perhaps more realistic.)

\section{Conclusions}

We showed how the existing (na\"{i}ve) partitioning approach to reducing ZX-diagrams contains much redundancy inherent among its calculations. We then demonstrated how this redundancy could be avoided with the use of parallelisation and by precomputing unique terms (the latter at the cost of an increased memory overhead). We thereafter introduced the notion of \textit{pairwise regrouping}, whereby the partitioned segments, after each being fully reduced independently, could be regrouped one pair at a time (always opting for the computationally cheapest option at each step). The culmination of these techniques is a smart partitioner-based method for strongly classically simulating quantum circuits, coded into an openly available Python package: the \textit{ZX-Partitioner} (available at \href{https://github.com/mjsutcliffe99/zxpartitioner}{https://github.com/mjsutcliffe99/zxpartitioner}).

We benchmarked our method versus both the direct decomposition (no partitioning) approach as well as the na\"{i}ve partitioning approach. Against both, we saw a runtime increase of orders of magnitude for certain sized random circuits and we demonstrated how its effectiveness varies with the interconnectedness of the circuit. Ultimately, we were able to show that even under realistically reasonable conditions, this method is able to outperform the alternatives by orders of magnitude.

There remains scope to further improve upon the techniques presented in this paper, with perhaps the most interesting area to explore being in optimising for the partitioning function itself. Specifically, the Clifford simplification routine for reducing the initial ZX-diagram could be rewritten to balance minimisation of edge connectivity as well as T-count, rather than exclusively the latter at the cost of the former. Appendix \ref{app:improvingPartitionability} covers some preliminary ideas along these lines and outlines the difficulties involved.

\nocite{*}
\bibliographystyle{eptcs}
\bibliography{smartpartitioning}

\appendix

\section{Efficient pairwise regrouping algorithm}
\label{app:cudacode}

In section \ref{sec:methods:regrouping}, we introduced the pairwise regrouping technique, with an implementation detailed in algorithm \ref{alg:regroupPseudo}. This is a very high-level algorithm, making use of high-level data structures (such as \textit{sets}) and functions (such as those pertaining to set theory and the handling of bitstrings). While we hope this makes the procedure relatively simple and straightforward to follow and helps convey how the method works, it also means this implementation is sub-optimal for the time-sensitive nature of its use case. Consequently, in listing \ref{listing:regroupGPU} we present a far more efficient, low-level implementation of this procedure, making use of binary encoding, bitwise calculations, and GPU parallelism.

\lstdefinestyle{myStyle}{
    belowcaptionskip=1\baselineskip,
    breaklines=true,
    frame=none,
    numbers=none, 
    basicstyle=\footnotesize\ttfamily,
    keywordstyle=\bfseries\color{green!40!black},
    commentstyle=\itshape\color{purple!40!black},
    identifierstyle=\color{blue},
    backgroundcolor=\color{gray!10!white},
}

\begin{lstlisting}[language=C, style=myStyle, label=listing:regroupGPU, caption={An efficient, low-level and GPU-parallelised CUDA kernel for pairwise regrouping.}]
__global__ void regroup_pair_gpu(int paramsA, int paramsB, int paramsC, float * A_re, float * A_im, float * B_re, float * B_im, float * AB_re, float * AB_im, const int N_params, const int size)
    {
        int index = blockIdx.x * blockDim.x + threadIdx.x;
        
        // LOCALLY INDEX...
        
        int ab = 0;
        int bc = 0;
        int ac = 0;
        int abc = index;
        int x = 0; // current length of ab
        int y = 0; // current length of bc
        int z = 0; // current length of ac
        
        for (int i=0; i<N_params; ++i)
        {
            if (paramsA & 1) ab = ((abc & 1) << x++) | ab;
            if (paramsB & 1) bc = ((abc & 1) << y++) | bc;
            if (paramsC & 1) ac = ((abc & 1) << z++) | ac;
            abc >>= 1;
            paramsA >>= 1;
            paramsB >>= 1;
            paramsC >>= 1;
        }
        
        // MULTIPLY SCALARS (A_ab * B_bc -> AB_abc) ...
        // (A+ai)(B+bi) = (AB-ab) + (Ab+aB)i
        
        float A = A_re[ab];
        float a = A_im[ab];
        float B = B_re[bc];
        float b = B_im[bc];

        atomicAdd(&AB_re[ac], (A*B) - (a*b));  //AB_re[index] = (A*B) - (a*b);
        atomicAdd(&AB_im[ac], (A*b) + (a*B));  //AB_im[index] = (A*b) + (a*B);
        __syncthreads();
    }
\end{lstlisting}

\lstset{language=C,style=myStyle}

This CUDA code shows the function \lstinline{regroup_pair_gpu}. Be aware that this is a CUDA kernel, rather than a conventional function, meaning it is executed many times in parallel upon the GPU threads. While this kernel code is C-like, it may be called from within Python, passing as arguments the local parameter sets of segments $A$ and $B$ respectively, together with their exclusively uncommon parameters (i.e. those of the future grouped $AB$ segment), and also the total number of parameters involved among $A$ and $B$.

Rather than using a high-level data structure like a \textit{set} or even a \textit{list}, we instead record sets of parameters as individual integers. In the example of table \ref{table:regroup}, we have segment $A$ containing parameters $\{a,b\}$ and segment $B$ containing parameters $\{b,c\}$. So, for a collective set of parameters $\{a,b,c\}$, we express this as \lstinline{paramsA = 110} and \lstinline{paramsB = 011}. (As integers, this would be interpreted as \lstinline{paramsA = 6} and \lstinline{paramsB = 3}, but for our purposes it makes more sense to interpret these as their binary bitstrings.) In this example case, we would also have \lstinline{paramsC = 101} being the exclusive uncommon parameter set (i.e. $\{a,c\}$, which will be the parameters in the upcoming regrouped segment, $AB$). Converting the sets into this form can be done quite trivially in Python in advance of calling the kernel, and then this data can be passed among its arguments, along with the total number of parameters involved (in this case, $A$ and $B$ collectively contain $3$ parameters, $\{a,b,c\}$, so \lstinline{N_params = 3}). Meanwhile, the argument \lstinline{size} is just the number of parallel threads to compute (i.e. the number of rows of table \ref{table:regroup}), $2^{N\_params}$. (Note that the remaining arguments of the kernel, such as \lstinline{A_re}, refer to the memory wherein the real and imaginary parts of the list of scalars of segments $A$ and $B$ are recorded, while those of \lstinline{AB_re} and \lstinline{AB_im} are initially just empty blocks of data, acting as empty arrays to which the kernel will be writing.)

The only difference among each parallel thread executing this kernel is its unique identifier number, \lstinline{index}, ranging from \lstinline{0} to \lstinline{size-1}. This \lstinline{index} essentially gives the \textit{values} of the parameters for this thread. For instance, in our example, \lstinline{index=5} - which in binary would be \lstinline{101} - would denote the case whereby $a=1,\; b=0,\; c=1$. (This is the \lstinline{(index+1)}\textsuperscript{th} row of table \ref{table:regroup}.) The logic of the \lstinline{for} loop in this kernel then serves to take this \lstinline{index}, representing the full set of parameters, $\{a,b,c\}$, and determine the respective \textit{local} sets of parameters of segments $A$ and $B$ - in this case $\{a,b\}$ and $\{b,c\}$. (So, for \lstinline{index=5}, we would take \lstinline{abc=101} and deduce \lstinline{ab=10} and \lstinline{bc=01}.) This works via clever usage of bitshifting and bitwise operations, and - for the keen reader - the best way to understand the logic is to work through the table \ref{table:regroup} example step-by-step.

For each (parallel) iteration of the kernel, one scalar, $AB_{ac}$, among the new regrouped segment will be calculated and saved to memory. The only remaining point to note here is the use of the \lstinline{atomicAdd(x,y)} CUDA function. This adds the value \lstinline{y} to the memory address \lstinline{x}, but does so in a parallel-friendly way which avoids race conditions \cite{cudaGuide}.

Regarding its use in the ZX-Partitioner, if the projected runtime is below a certain threshold then the high-level implementation of algorithm \ref{alg:regroupPseudo} is used (coded in Python), as on such low scales the overhead in initialising the data to the kernel makes the GPU approach actually slower than the higher-level implementation. However, for sufficiently (non-trivially) sized cases, the efficient CUDA implementation is used, giving a drastic performance speedup (for this particular part of the computations).

\section{Projected runtimes}
\label{app:projectedRuntimes}

It is imperative that many of the experiments in this paper observe the results at very large scales and, due to their large variance, require lots of repeats. Unfortunately, taking many hundreds of such measurements - each taking potentially many hours or beyond - would be prohibitively impractical. Consequently, for the results in this paper, we opted instead to record the \textit{projected} (or \textit{estimated}) \textit{runtimes}. This allows us to determine reasonable runtime measurements without having to actually execute the (potentially very slow) classical simulation methods.

In each case, we can straightforwardly calculate the number of calculations (of each type) that each method would perform. We then need only divide this by the pre-measured average rates at which these calculations may be computed. For instance, the direct decomposition (no partitioning) approach requires calculating $S_{decomp} = 2^{\alpha t}$ stabiliser terms, for a given initial (post Clifford simplification) T-count $t$. Moreover, this paper use the state-of-the-art \textit{cats} family of decompositions \cite{kissinger2022simulating2}, which, in practise (and allowing for inter-step simplification), we observed to have an average $\alpha\approx0.32\pm0.02$. As such we take $\alpha=0.32$ when calculating our $S_{decomp}$ readings. Similarly, the number of precomputing calculations, $S_{precomp}$, and cross-referencing calculations, $S_{crossref}$, can be respectively determined for the smart partitioner approach as outlined in section \ref{sec:methods:regrouping}.

To determine the average number of calculations per second that each calculation type achieves, we conducted over a hundred measurements for (non-trivially sized) circuits ranging in depth, qubit count, T-count, and partitionability. In each case, we executed the different simulation methods and recorded the resulting runtimes. (For the smart partitioner method, this was broken down into the precomputing part and the cross-referencing part.) From these experiments, we observed the following averaged runtime rates (with standard deviation error margins), measured in \textit{calculations per second} to $3$ significant figures:

\[ R_{decomp} = 1,730 \pm 650 \;\;calcs/s \]
\[ R_{precomp} = 21,400 \pm 13,300 \;\;calcs/s \]
\[ R_{crossref} = 412,000 \pm 145,000 \;\;calcs/s \]

These measurements were recorded on a commercial laptop with an \textit{11\textsuperscript{th} Gen Intel Core i5-11400H @ 2.70GHz} CPU, \textit{NVIDIA GeForce GTX 1650} GPU, and \textit{8GB SODIMM} RAM. Given these rates, we can estimate the runtime of the direct decomposition (i.e. no partitioning) approach, as well as the smart partitioner approach, as:

\[ T_{decomp} = \frac{S_{decomp}}{R_{decomp}} \]

\[ T_{smart} = T_{overhead} + \frac{S_{precomp}}{R_{precomp}} + \frac{S_{crossref}}{R_{crossref}} \]

The overhead time from the partitioning function itself, $T_{overhead}$, varied with the size and shape of the circuit, but was never beyond a few seconds (and typically below a second) and hence, beyond the trivially quick cases, provided a negligible contribution to the overall runtime. Regardless, unlike the precomputing and cross-referencing times, this was measured (rather than estimated) in every case, so that when it provided a non-negligible contribution to the total runtime this is appropriately reflected in the recorded results.

It is understandable that the cross-referencing calculations can be computed much more rapidly than decomposing and precomputing as these calculations are very simple, as shown in section \ref{sec:methods:regrouping}, and can be processed efficiently with GPU parallelism (see appendix \ref{app:cudacode}). On the other hand, each `calculation' in decomposing involves many steps of ZX-diagram simplification. The precomputing calculations, meanwhile, are essentially equivalent to those of decomposing except in that (as emphasised in section \ref{sec:methods:paramzx}) they too have space for GPU parallelism \cite{sutcliffe2024fastclassicalsimulationquantum}. Lastly, note that the decomposing calculations (within also the precomputing calculations) are computed via Quizx \cite{github:quizx} rather than Pyzx, to ensure this is as speedy as possible.

Estimating the runtimes in this way is justified in that the results we show in this paper aim to highlight \textit{scales} and \textit{trends}, rather than exact numerical runtimes (which at any rate would vary with hardware). Notably, these runtime rates are rather consistent (certainly with regard to order of magnitude). Indeed, the results we present in this paper are plotted on logarithmic scales and so the variance due to the uncertainty in the runtime rates above would not make a noticeable difference. This is especially true given that these small uncertainties would be negligible compared to the existing magnitudes-wide error margins (i.e. in figure \ref{fig:timevsigmaCapped}) due to the variance among the different ZX-diagrams.

As each cross-reference calculation can be computed much more quickly than each precomputation calculation, the program can in fact aim to balance the projected \textit{runtimes} of these two parts, rather than simply their number of calculations. This step improves efficiency by ensuring neither part dominates the overall runtime. Lastly, note that, in all runtime results shown in this paper, we quote the projected runtime \textit{unless} this is below $100$ seconds, in which case we compute and measure the real runtime. This ensures that speedy results - where there is more fluctuation and minor contributions to runtime can no longer be assumed to be negligible - are also accurate.

As a brief note on the memory overhead for the ZX-Partitioner, this scales as $\max{(\minpair{(H_i)})}$ for $i\in\{0,1,\ldots,k-2\}$ (see section \ref{sec:methods:regrouping}). What this means is that, while this can result in Gigabytes of memory overhead, at scales beyond this the runtime would already become infeasible. In other words, assuming Gigabytes to be the upper-bound of what would be feasible, the memory overhead tends not to become infeasible before the runtime would - so this is not a limiting factor for the method.

\section{Additional results and figures}
\label{app:additionalResults}

\begin{figure}
    \centering
    \includegraphics[scale=0.54]{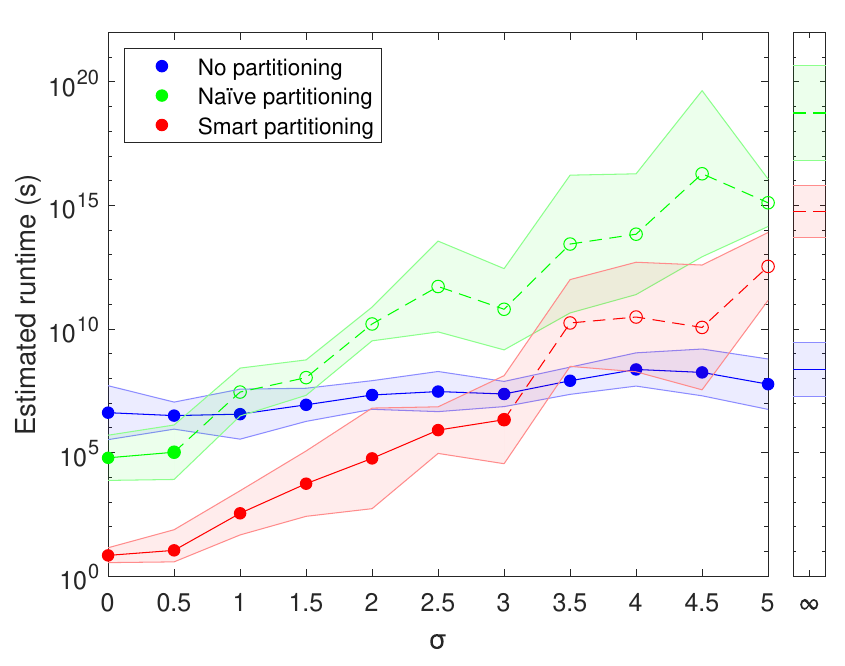}
    \caption{The average projected runtimes for strongly simulating Clifford+T circuits of $30$ qubits and a depth (gate count) of $1,000$, using three different methods. This varies from figure \ref{fig:timevsigmaCapped} in that here we enforce that partitioner methods always make at least one partition ($k\geq2$).}
    \label{fig:timevsigmaUncapped}
\end{figure}

In figure \ref{fig:timevsigmaCapped}, we considered strong simulation of quantum circuits of a particular size ($1,000$ gates across $30$ qubits) using the novel smart partitioning method outlined in this paper, as well as the existing partitioning method and the baseline decomposition approach. The figure shows how the variance, $\sigma$, of CNOT spread across the qubits impacted this runtime. One notable observation from this figure is that neither partitioner method ever performs noticeably worse than the direct decomposition approach. This is due to the fact that \textit{not} partitioning is essentially a special case of partitioning (with the most optimal $k=1$). While this is true, it can be both interesting and insightful to consider how this figure would change if the partitioner methods enforced always \textit{at least one partition} (i.e. $k\geq2$). This modified case is shown here in figure \ref{fig:timevsigmaUncapped}.

Here we can see that by always enforcing the partitioner methods to partition, they are liable (where applicable) to perform far worse than simply leaving the graphs unpartitioned. This further emphasises that, beyond a certain $\sigma$ threshold for certain sizes of circuit, the most optimal number of parts into which the diagram should be partitioned is indeed $k=1$. To this end, the more interconnected the graph is, the less likely it appears that larger $k$ partitions will prove profitable.

\begin{figure*}
    \centering
    \begin{subfigure}[t]{0.5\textwidth}
        \centering
        \includegraphics[scale=0.54]{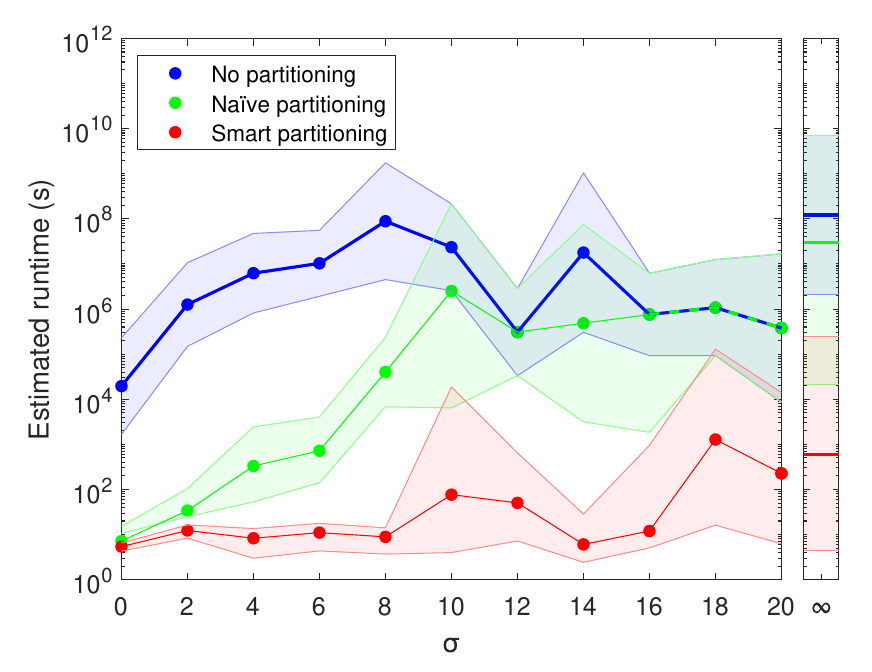}
    \end{subfigure}%
    ~ 
    \begin{subfigure}[t]{0.5\textwidth}
        \centering
        \includegraphics[scale=0.54]{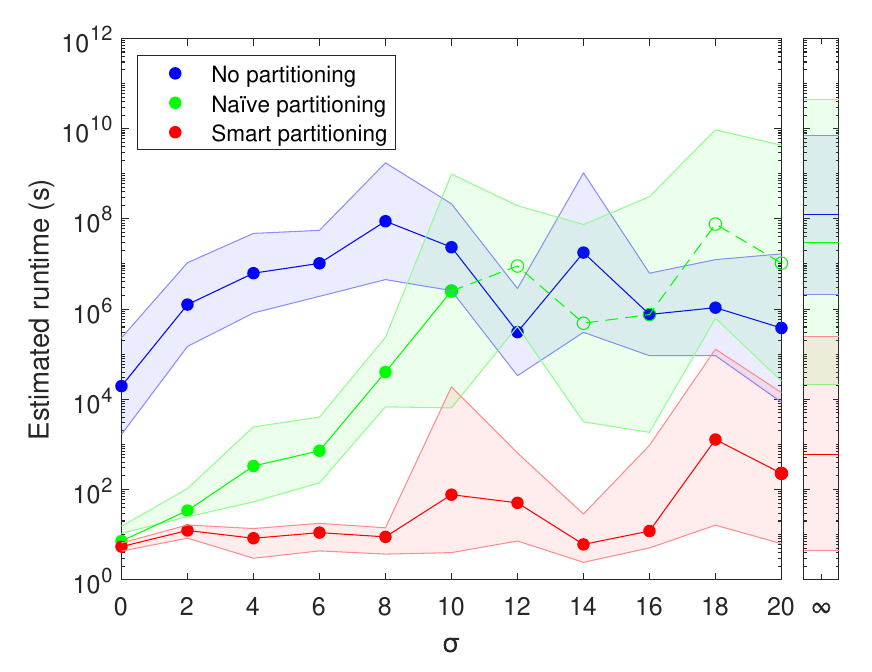}
    \end{subfigure}
    \caption{The average projected runtimes for strongly simulating Clifford+T circuits of $110$ qubits and a depth (gate count) of $1,000$, using three different methods. On the left, we allow the partitioner methods to make no partitions ($k=1$) where it deems appropriate, while on the right we enforce that they always make at least one partition ($k\geq2$).}
    \label{fig:timevsigmaB}
\end{figure*}

In figure \ref{fig:timevsigmaB}, we show a second example of how the runtime varies with $\sigma$ - this time for circuits of $1,000$ gates and $110$ qubits. For circuits of this size, as we already saw in figure \ref{fig:heatInf}, the smart partitioner approach always outperforms the na\"{i}ve approach, regardless of $\sigma$, as partitioning is always better than not. Even as $\sigma\rightarrow\infty$, the terminal runtime for smart partitioning is $T_{smart}^{\sigma\rightarrow\infty} = 10^{2.77\pm2.69}$ seconds, while for direct decomposition is $T_{direct}^{\sigma\rightarrow\infty} = 10^{8.03\pm1.93}$ seconds.

\begin{figure*}[t!]
    \centering
    \begin{subfigure}[t]{0.5\textwidth}
        \centering
        \includegraphics[scale=0.7]{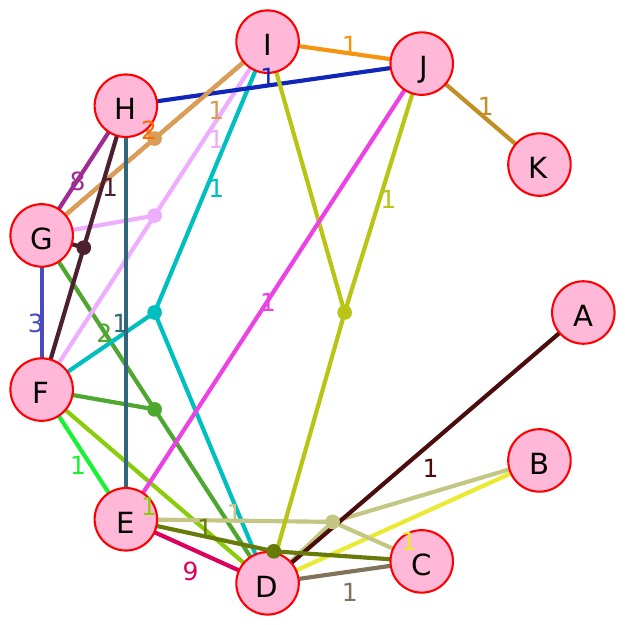}
    \end{subfigure}%
    ~ 
    \begin{subfigure}[t]{0.5\textwidth}
        \centering
        \includegraphics[scale=0.7]{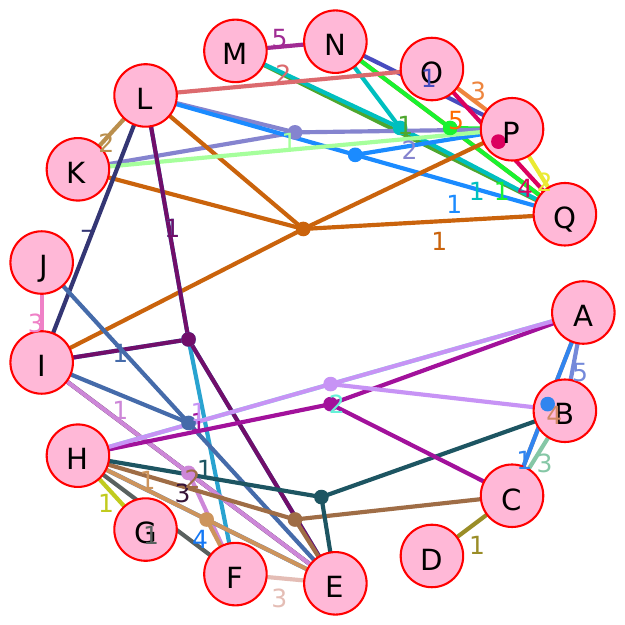}
    \end{subfigure}
    \caption{Two heftier examples of segment connectivity hypergraphs, generated from partitioning random ZX-diagrams via the ZX-Partitioner.}
    \label{fig:bigHnets}
\end{figure*}

In figure \ref{fig:hNet0}, we showed a hypergraph (generated by the ZX-Partitioner) denoting the connectivity among partitioned segments for a particular example case. This was a fairly simple example, for illustrative purposes, so we also present - in figure \ref{fig:bigHnets} - two heftier examples.

\section{Tensor contraction and compound circuits}
\label{app:compoundCircuits}

With regard to its scope and computational complexity, the smart partitioner method detailed in this paper appears relatively similar to the tensor contraction \cite{markov2008simulating,brennan2021tensor} approach to strong classical simulation. Both have a memory overhead and runtime complexity that grow exponentially with the interconnectedness (or \textit{treewidth}) of the circuit. Indeed, for the experiments run in this paper, many of the cases that were particularly favourable to the smart partitioner approach (such as shallow circuits or those with especially low $\sigma$) were also effectively simulable with the tensor contraction method.

Despite this, it is not strictly true that all circuits for which the smart partitioner method is effective could also be effectively simulated via tensor contraction. In fact, it is very easy to design example cases which showcase this point. As a very simple example, consider a set of subgraphs, $G_1,\ldots,G_5$, which are each \textit{internally} very highly interconnected such that each individually is beyond the scope of tensor contraction but within the scope of stabiliser state decomposition. Now suppose these subgraphs are connected \textit{to one another} in a relatively inexpensive way (that is, with relatively few edges).

As each subgraph is individually beyond the scope of tensor contraction, it follows that whole is likewise. Nevertheless, the smart partitioner method could very effectively (and at a relatively small computational cost) partition the graph into its $5$ locally-dense subgraphs and reduce each using stabiliser decomposition, before cross-referencing the results to attain the final amplitude. Indeed, on randomly generated examples similar to this, we were able to verify that tensor contraction (using the \textit{Quimb} \cite{gray2018quimb} Python library) would fail (due to exceeding a reasonable runtime limit or 128GB of memory overhead) while the smart partitioner would complete within seconds and requiring (in some cases) only a matter of bytes in memory overhead.

So, while there was a notable overlap in the applicability of these two methods on the types of circuits used in the experiments presented in this paper, this essentially is a consequence of the means by which the random circuits were generated. With slight modification, we can generate a similar class of random circuits, which have non-uniform CNOT spreads and are realistically justified and which are (generally) effectively simulated via the smart partitioner but not tensor contraction.

Specifically, we can randomly generate $k$ distinct Clifford+T circuits of $q$ qubits (with uniform CNOT spreads, i.e. $\sigma=\infty$). These circuits may be vertically composed and some number, $n$, of additional CNOTs may be inserted which each connect \textit{between} some pair of the sub-circuits. For each of these external CNOTs, when deciding which how far away the target sub-circuit should be from the source, we can - as before - use a normal distribution (albeit acting on a sub-circuit by sub-circuit level rather than a qubit by qubit level). Generating circuits in this way leads to structures like the following:

\ctikzfig{compound_circuit}

These \textit{`compound circuits'} manifest highly interconnected local cliques, which in turn are connected to one another by only a relatively modest number of edges. Moreover, they arguably offer fairly realistic examples of circuit structures, being composed of smaller independent subroutines which relay some information to one another.

While it is difficult to fairly quantify such results (as such circuits can be made as generously or ungenerously to our aims as desired), in preliminary experiments, we observed that - generally speaking - such circuits are practically unsimulable for both tensor contraction and direct decomposition, yet are effectively simulated with smart partitioning.

\section{Improving partitionability}
\label{app:improvingPartitionability}

While this paper presents a number of optimisations to the calculations involved in a partitioning-based method of quantum circuit simulation, the partitioning function itself remains essentially untouched from graph theory literature. Significantly, the partitioner treats ZX-diagrams as generic graphs and does not know anything of how the diagrams may be transformed via the rewriting rules. Rectifying this issue, by taking into account ZX-calculus based optimisations for improving graph partitionability, could yield potentially drastic further reductions to the runtime. Indeed, this could be a very interesting area of research (and one extensive enough to warrant its own paper, beyond the scope of this one), being a twist on the usual simplification strategies which aim to exclusively minimise, for instance, T-count (usually at the expense of increased edge connectivity).

To briefly show a proof of concept of the kinds of considerations one could make, examine the following rewriting rules (known respectively as \textit{local complementation} and \textit{pivoting} \cite{kissinger2022simulating1}):

\ctikzfig{deriv_rules}

These rules (derived from the basic set of figure \ref{fig:zxrules}) are among the most important and widely used in scalar diagram reduction. Generally, these rules, whenever applicable, are applied left to right to aid in spider (and hence T-count) minimisation. However, notice that this comes at the cost of significantly increasing the connectivity among the remaining spiders, which is liable to considerably hinder partitionability. Thus, it might be advisable to be more discriminatory when deciding whether to apply these rules, or even to apply them in reverse (right to left) where it might be appropriate. For instance, in many cases it will likely be worthwhile to \textit{un-gadgetise} the phase gadgets (see \cite{kissinger2022simulating1}) after full Clifford reduction.

Likewise, there exists the \textit{bialgebra} rule \cite{van2020zx}:

\ctikzfig{bialgebra}

Appropriately applying this rule in reverse (right to left) could also be very helpful in aiding partitioning. Suppose the leftward edges connected to some subgraph, $G_A$, and the rightward edges to some other subgraph, $G_B$, which are otherwise unconnected. By applying the bialgebra rule in reverse one can minimise the number of cuts required to separate these two subgraphs from $\texttt{min}(n,m)$ down to just $1$. Perhaps even more helpfully, if each of the $n+m$ outgoing edges of this diagram were connected to its own unique (and otherwise disconnected) subgraph, then by reversing bialgebra the number of cuts required to fully disconnect all these subgraphs would be brought from $\texttt{min}(n,m)$ down to just $2$.

Moreover, from local complementation and the cutting decomposition (plus the known rule whereby two parallel Hadamard edges between a pair of spiders may cancel out \cite{van2020zx}), we may derive a new rule which allows us to toggle the (Hadamard) edge connectivity among any set of $n\geq2$ like-coloured spiders at the cost of one cut, as in the following example:

\ctikzfig{toggleConnectivity}

Notice that in the initial diagram (left-hand side) of the above example, we could alternatively have chosen to toggle the connectivity among the set $\{\alpha_1,\alpha_2,\alpha_3\}$ or $\{\alpha_1,\alpha_2,\alpha_4\}$, or even $\{\alpha_1,\alpha_4\}$ or $\{\alpha_2,\alpha_3\}$, etc. Each of these options would have represented fully connected cliques and so applying our new rule in any of these cases would have only removed edges and not introduced any new ones. Nevertheless, the best of these options would have only partitioned the diagram into $2$ disconnected parts, whereas the incomplete (but near) clique set of $\{\alpha_1,\alpha_2,\alpha_3,\alpha_4\}$, which we chose to apply the rule to, enabled partitioning into $3$ disconnected parts.

When deciding when and where to apply this rule, it is not always obvious, particularly among larger graphs with lots of connected cliques and near-cliques, what the best set of spiders to select in each case is. The following example illustrates this point:

\ctikzfig{reverseLC}

Here, we have $5$ otherwise disconnected subgraphs, $G_1,\ldots,G_5$, that meet among these $5$ spiders, $\alpha_1,\ldots,\alpha_5$. There are many different ways in which we can apply the above rule to fully disconnect these subgraphs, with one option for the cheapest approach (costing just $2$ cuts) shown here. Yet, devising some algorithm to determine the optimal applications of this rule (particularly for cases too large and complex to deduce by inspection) is something which remains for further research.

\end{document}